\documentclass{article}
 \usepackage[utf8]{inputenc}
 \usepackage[english]{babel}

 \usepackage{natbib}
 \usepackage{apalike}
\usepackage{times}

 \usepackage[hidelinks]{hyperref}
\usepackage[]{graphicx}
\chardef\bslash=`\\ 

\usepackage{authblk}
\usepackage[labelfont=bf,labelsep=space]{caption}
\usepackage{tabularx}
\usepackage{multirow}
\usepackage{booktabs}
\usepackage{url}
\usepackage{enumitem}
\newlist{tabitemize}{itemize}{1}
\setlist[tabitemize]{label=\textbullet,leftmargin=*,topsep=1ex,
parsep=0pt,
                  after=\vspace{-\baselineskip},
                  before=\vspace{-0.75\baselineskip}
                  }  
\newcolumntype{L}[1]{>{\raggedright\arraybackslash}p{#1}}
\usepackage{geometry}
\usepackage{array}
\usepackage{multirow}
\usepackage{tabularx}
\usepackage{makecell}
\usepackage{url}
\DeclareRobustCommand{\rchi}{{\mathpalette\irchi\relax}}
\newcommand{\irchi}[2]{\raisebox{\depth}{$#1\chi$}} 
\usepackage[dvipsnames]{xcolor}
\usepackage{xspace}
\newcommand{\ngu}{N19\xspace}
\newcommand{\rapp}{R19\xspace}
\newcommand{\zhou}{Z21\xspace}
\newcommand{\osabe}{O21\xspace}
\newcommand{\pinsplus}{PINSPlus\xspace}
\newcommand{\nemo}{NEMO\xspace}
\newcommand{\sfmeb}{SFMEB\xspace}
\newcommand{\mbcdeg}{MBCdeg\xspace}
\newcommand{\mbcdega}{MBCdeg1\xspace}
\newcommand{\mbcdegb}{MBCdeg2\xspace}
\newcommand{\iclusterplus}{iCluster+\xspace}
\usepackage{abstract}

\begin{document}

\title{Explaining the optimistic performance evaluation of newly proposed methods: a cross-design validation experiment}
\author[1]{Christina Nie{\ss}l\thanks{\sf{e-mail: cniessl@ibe.med.uni-muenchen.de}}}
\author[1,2]{Sabine Hoffmann}
\author[1]{Theresa Ullmann}
\author[1]{Anne-Laure Boulesteix}
\affil[1]{Institute for Medical Information Processing, Biometry and Epidemiology, LMU Munich, Marchioninistr. 15, 81377, München, Germany}
\affil[2]{Department of Statistics, LMU Munich, Ludwigstr. 33, 80539, München, Germany}

\maketitle 

{\begin{abstract} \normalsize
The constant development of new data analysis methods in many fields of research is accompanied by an increasing awareness that these new methods often perform better in their introductory paper than in subsequent comparison studies conducted by other researchers. We attempt to explain this discrepancy by conducting a systematic experiment that we call ``cross-design validation of methods’'.  In the experiment, we select two methods designed for the same data analysis task, reproduce the results shown in each paper, and then re-evaluate each method based on the study design (i.e., data sets, competing methods, and evaluation criteria) that was used to show the abilities of the other method. We conduct the experiment for two data analysis tasks, namely cancer subtyping using multi-omic data and differential gene expression analysis. Three of the four methods included in the experiment indeed perform worse when they are evaluated on the new study design, which is mainly caused by the different data sets. Apart from illustrating the many degrees of freedom existing in the assessment of a method and their effect on its performance, our experiment suggests that the performance discrepancies between original and subsequent papers may not only be caused by the non-neutrality of the authors proposing the new method but also by differences regarding the level of expertise and field of application.
\end{abstract}}
\newpage
\section{Introduction}
In the literature on data analysis methods, including statistical journals, machine learning journals, and conference proceedings, most articles describe {\it new} methods, thus contributing to an increasing multitude of potential methods addressing various data analysis problems. It is commonly claimed by the authors proposing these new methods that they perform better than existing ones in some sense. For anecdotal evidence in the context of supervised classification see \cite{boulesteix2013plea}. 
The fact that new methods are typically claimed to be better than existing ones does not necessarily imply that these statements are wrong. In fact, this is what we would expect if we assume continuous scientific progress. However, the recurrent character of these claims, combined with the requirement of journals and reviewers to make these sort of claims regarding the superiority of the proposed methods, make them somewhat suspicious \citep{norel2011self,boulesteix2013plea}. 
A recent survey of papers that compare pre-processing methods for a special type of high-throughput molecular data indicate that, at least in this specific context, the paper introducing a method is indeed more optimistic regarding its performance than subsequent papers that are more neutral towards the method \citep{buchka2021optimistic}. 

In a different but related approach, several studies demonstrate that it is relatively easy to make a method appear better than it actually is  \citep{jelizarow2010over,ullmann2022over,pawel2022pitfalls,Niessl2021,sonabend2022avoiding,Boulesteix2009}. These studies suggest that over-optimistic statements regarding a method's performance may be partly attributed to the non-neutral attitude of the authors, who are naturally interested to present their method in a positive light. More precisely, it is argued that the non-neutrality may translate into a conscious or subconscious optimization of the method and the study design in which it is evaluated (e.g., by selectively reporting the considered data sets or simulation parameters) such that the proposed method shows good performance.

Imagine there are two methods, method A and method B, available to address a specific data analysis task. We set the number of methods to two for the sake of simplicity, but the following arguments can be extended to a setting with more than two methods. Further, imagine the typical situation in which the authors of A and the authors of B both claim that their method performs well. The study designs they use to support their claims are different. We will call them design A and design B, respectively. 
Following the conjecture discussed in the previous paragraph that study designs used by authors of methods may overfit their methods and vice-versa, a natural question is how method A would perform when re-evaluated using design B and how method B would perform when re-evaluated using design A.  
In the present paper, we put this idea into practice by conducting a systematic experiment, which we call \lq\lq cross-design validation of methods''. More precisely, we consider two exemplary data analysis tasks, namely multi-omic data integration for cancer subtyping and differential gene expression analysis, and for each exemplary task, we select two papers that propose a new method. For each of these two methods, we reproduce the evaluation shown in the paper that introduced it, and then re-evaluate it on the design used by the authors of the other paper. In this context of {\it methodological} research, where data analysis methods are considered as research objects, the study design includes data sets (with a focus on real data for the first task and simulated data for the second task), competing methods, and evaluation criteria. 

The goal of this cross-design validation experiment is two-fold. Firstly, it  allows dissection of the variability of study designs and its impact on the results in the context of methodological research---in a similar way as so-called multi-analyst experiments \citep{silberzahn2018many} do in application fields of statistics. Secondly, the cross-design validation experiment provides insights into the mechanisms leading to performance discrepancies such as those observed by \cite{buchka2021optimistic}  between the original paper (i.e., the paper that introduces the method of interest) and subsequent papers (i.e., papers that propose another method and include the method of interest as competitor or papers that are dedicated to method comparison itself). Importantly, these are {\it real-world} observations---as opposed to the previous experiments by \cite{jelizarow2010over}, \cite{ullmann2022over}, and \cite{pawel2022pitfalls}, mimicking the behavior of fictional researchers who wish to present their method in a favorable light.
Finally, our experiment also provides insights regarding the reproducibility of results and the difficulty of performing fair method comparisons as a by-product. 

Because the authors of our selected papers made code and data available for the purpose of reproducibility, our experiment can be performed without involving them personally, which considerably simplifies its organization and execution. Moreover, while we could also gain insights from re-evaluating the methods on a study design selected by ourselves, the cross-design character of the experiment guarantees a certain degree of neutrality of our comparisons.  

 The remainder of this paper is structured as follows. The general structure of our cross-design validation experiment is outlined in Section~\ref{section:experiment}. Section~\ref{section:omics} and \ref{section:de} present the two data analysis tasks, while a discussion of the mechanisms leading to the observed performance differences along with a summary can be found in Section~\ref{section:discussion}. We conclude the paper in Section~\ref{section:conclusion}.

\section{Preliminary remarks and design of the experiment}\label{section:experiment}

\subsection{Terminology}\label{section:experiment_term}
Before describing the experiment in more detail, we briefly clarify the terminology used throughout the paper. Similar to  \cite{Klau2020} and \cite{buchka2021optimistic}, we define the term \textit{method} not just as the  statistical testing or modeling approach, but the full analysis pipeline, potentially including steps such as data normalization. All methods considered in the experiment have several parameters that can be set by the method user (e.g., the maximum number of clusters or the type of multiple testing correction), which we refer to as \textit{method parameters}. 

Moreover, we define the \textit{study design} as the combination of all components that contribute to the performance assessment of the method of interest. The study design consists of three main components, 
namely data sets (real or simulated), competing methods (including their respective method parameters), and evaluation criteria (in our exemplary analysis tasks referring to the evaluation metric and the way the results are aggregated across the real data sets or simulation repetitions). Note that data pre-processing can be seen both as part of the method or part of the data component. In our experiment, we consider pre-processing steps as belonging to the data component if performed for \textit{all} methods and belonging to the method (i.e., the method of interest or the competing methods) otherwise.

\subsection{Selection of the papers}
As a preliminary step, we first have to select appropriate papers for both exemplary data analysis tasks that we consider in our experiment, namely cancer subtyping using multi-omic data and differential gene expression analysis. Both are applications from the field of biostatistics at the interface with bioinformatics. Apart from the requirement that the paper must introduce a new method, there are two eligibility criteria related to reproducibility: (i) the code to reproduce the results presented in the paper is \textit{publicly} available and can be run without errors, and (ii) the code is written in \texttt{R}. 

While the restriction to \texttt{R} as programming language (ii) excludes some papers, the majority of papers fail criterion (i). In many cases, authors only provide the code to use their method (e.g., an \texttt{R} package) but not to reproduce the results shown in the paper. In other cases, the link to the code is broken, the code supposedly included in the supplement cannot be found, or some of the files needed to reproduce the code are missing (e.g., the file containing the empirical data the simulation shown in the paper is based on). Note that we purposely refrain from contacting the authors if the code is not publicly available to make the selection of papers independent of the authors' willingness to respond and provide the code. 
Although we do not restrict the number of papers to two per data analysis task in advance, the above-mentioned difficulties lead us to stop the search after finding two eligible papers, resulting in  $2\times 2=4$ papers included in our experiment.

The conclusion of this informal search process is that the practice of making code and data openly available is far from being the standard in the methodological literature beyond positive exceptions such as the Biometrical Journal \citep{Hofner2016}.
The four papers included in our experiment \citep{Nguyen2019, Rappoport2019,Zhou2021,Osabe2021} should thus be seen as rare positive examples for open research practices in methodological research.

\subsection{Design of the experiment}\label{section:experiment_design}
While all four papers evaluate their respective method extensively in various settings, our experiment includes only the results that (i) are presented as figures or tables and appear in the main paper, i.e., excluding the supplement (to keep the experiment feasible) and (ii) compare the method of interest to competing methods (since we can only compare the \textit{relative} performance of a method if the considered papers use different evaluation metrics that do not allow a direct comparison). If the results are based on both real and simulated data, we only consider the results of the data type that is predominantly employed in the paper. In some cases, we exclude more results, which is reported and justified in Section~\ref{section:design_omics} and~\ref{section:design_de}. 

For each of the four papers, we first compare the results obtained by running the available code to the results presented in the corresponding paper. We use the same \texttt{R} and \texttt{R package} versions and do not modify the code in a way that would change the results, even in cases where we notice discrepancies between the code and the procedure described in the paper (referred to as ``design-implementation-gap'' by \citealp{lohmann2021s} in the context of simulation studies). Exceptions to this rule are explicitly reported in Section~\ref{section:challenge_omics} and~\ref{section:challenge_de}.

For both data analysis tasks, we then re-evaluate each method on the study design used by the authors of the other paper and compare the resulting performances. Our experiment can thus be seen as ``cross-design validation of methods'' (see Table~\ref{tab:experiment}). 
As stated above, the study design consists of three main components, namely data sets, competing methods, and evaluation criteria. We also vary these components individually, which allows us to assess their individual impact on the performance of the selected methods. Some challenges arise when
re-evaluating the methods on the new study design, in particular the choice of method parameters, which we set before viewing the performance results to avoid the risk of favoring one of the methods. Moreover, while we generally adhere to the code used to reproduce the results when ``crossing'' the designs, some modifications are necessary. Details on how we address these challenges for each data analysis task can be found in Section~\ref{section:challenge_omics} and~\ref{section:challenge_de}.

The \texttt{R} code and data to reproduce the experiment are openly available at \url{https://doi.org/10.6084/m9.figshare.20754028.v1}

\begin{table}[t]
\caption{Illustration of the cross-design validation experiment}
\label{tab:experiment}
\begin{tabular}{| >{\raggedright\arraybackslash}m{3cm}| >{\centering\arraybackslash}m{4cm}| >{\centering\arraybackslash}m{5cm}|  } 
\hline
  & Performance of method~$A$ & Performance of method~$B$ \\  \hline
 Study design by authors of method $A$ & Shown in original paper & ?\\  \hline
 Study design by authors of method $B$ & ? &  Shown in original paper    \\ 
 \hline
\end{tabular}
\end{table}

\section{Data analysis task I: Cancer subtyping using multi-omic data}\label{section:omics}
The first exemplary data task we consider in our experiment is cancer subtyping through clustering of patients based on multi-omic data, an active research field with many newly proposed methods in recent years (see \citealp{Duan2021} for an overview). The aim of these methods is to identify clusters (in this context referred to as \textit{subtypes}) with common biological characteristics or clinical phenotypes (e.g., survival time or drug response). This process helps to understand  the etiology of the disease and to develop better diagnostic tools and personalized treatment strategies \citep{Subramanian2020,Duan2021,Tepeli2020}. Recently developed cancer subtyping methods are usually able to integrate multiple types of high-dimensional molecular data such as genomics, epigenomics, transcriptomics, or proteomics (hence the term \textit{multi-omic} data; \citealp{Subramanian2020}).
 The two methods selected for our experiment are PINSPlus  and NEMO, which were proposed by \cite{Nguyen2019} and \cite{Rappoport2019}, respectively. Information on where to find the original codes provided by the authors is listed in our code documentation. We will abbreviate \cite{Nguyen2019} and \cite{Rappoport2019} by \ngu and \rapp.

\subsection{Study design in the original papers}\label{section:design_omics}
In the following, we outline and compare the study designs that are used to assess the performance of \pinsplus and \nemo in their respective original papers and that meet the inclusion criteria of our experiment (see Table~\ref{tab:design_multiomics} for an overview). We also report the authors' justifications of the design choices. For this purpose, we will also refer to  \cite{Nguyen2017}, which propose PINS, the predecessor method of PINSPlus, and to \cite{Rappoport2018}, a benchmark study intended as neutral that has been previously conducted by the authors of NEMO. All results of NEMO's competing methods originate from this benchmark study, i.e., the results of NEMO were simply added to the results of the previously published benchmark study. Since both \rapp and \ngu mainly use real data sets to evaluate their methods,  we do not further consider the simulation study presented by \rapp.  \\

\begin{table}[ht!] \footnotesize
\caption{Overview of the study design components used for performance assessment of PINSPlus and NEMO. Included are only components (i) for which the corresponding results are presented as figures or tables in the main paper (i.e., not in the supplement), (ii) that compare the method of interest to other competing methods, and (iii) that correspond to the performance assessment based on real data.  In addition, some components are not included in the experiment, which are indicated by asterisks (*). Competing methods and evaluation criteria for data sets not included in the experiment are not shown.}
\label{tab:design_multiomics}
\begin{tabularx}{\textwidth}{|X|L{2cm}|L{5cm}|L{5cm}|}
 \hline  
 \textbf{Study design component} & & \textbf{PINSPlus} \citep{Nguyen2019} & \textbf{NEMO} \citep{Rappoport2019}\\ 
\hline
\multirow{2}{4em}{\textbf{Data sets}}  & Number and type
&
\begin{tabitemize}
\item 34 TCGA data sets (gene expression, methylation, miRNA expression)
\item	*2 METABRIC data sets (gene expression, CNV)
\end{tabitemize} &
\begin{tabitemize}
\item 	10 TCGA data sets (gene expression, methylation and miRNA expression) 
\item *Partial TCGA data sets
\end{tabitemize}\\
\cline{2-4}
&
Pre-processing (all methods)  &
See Table \ref{tab:design_multiomics_supp} &
See Table \ref{tab:design_multiomics_supp} \\ 
\hline
\multirow{3}{4em}{\textbf{Competing \newline methods}}& 
Number and type & 
3: SNF, \iclusterplus, Consensus Clustering  &
9: PINS, SNF, iClusterBayes, $k$-means, spectral clustering, MCCA, LRACluster, *rMKL-LPP, *MultiNMF\\
\cline{2-4}
&
Pre-processing (method-specific)  &
See Table \ref{tab:design_multiomics_supp} &
See Table \ref{tab:design_multiomics_supp} \\
\cline{2-4}

&
Other method parameters &
\begin{tabitemize}
\item SNF: \textit{alpha} = 0.5,
 \textit{no. iterations} = 10,
\textit{no. clusters} = estimated according to eigen-gaps,
 \textit{max. no. clusters} = 5
\textit{no. neighbors} = 20
\item Others methods: see original paper
\end{tabitemize} & 
\begin{tabitemize}
\item SNF: \textit{alpha} = 0.5,
\textit{no.  iterations} = 30,
 \textit{no.  clusters} = estimated according to rotation cost,
 \textit{max. no. clusters} = 15,
 \textit{no. neighbors} = no.  of samples/10
\item Others methods: see original paper
\end{tabitemize} 
\\
\hline
\multirow{2}{4em}{\textbf{Evaluation criteria}}  &
Metric &
\begin{tabitemize}
\item 	Survival: logrank test 
\end{tabitemize} &
\begin{tabitemize}
\item 	Survival: permutation-based logrank test
\item Clinical: permutation-based $\rchi^2$/Kruskal Wallis test (discrete/continuous) for up to six clinical variables
\item 	*Runtime
\item	*Number of clusters
\end{tabitemize} \\ \cline{2-4}
&
Aggregation &
\begin{tabitemize}
\item	Number of data sets with significant and most significant logrank $p$-value
\end{tabitemize} &
\begin{tabitemize}
\item	Number of data sets with significant logrank $p$-value
\item	Number of data sets with at least one enriched clinical variable
\item	Mean  $-\mathrm{log}_{10}$ logrank $p$-value
\item	Mean number of enriched clinical variables

\end{tabitemize} \\ \hline
\end{tabularx}
\end{table}

\noindent\textbf{Data}\space\space 
Both \rapp and \ngu use data sets from The Cancer Genome Atlas Research Network (TCGA;  {\url{https://www.cancer.gov/tcga}}), where each data set corresponds to a different cancer type (e.g., kidney renal clear cell carcinoma or acute myeloid leukemia). The two author teams also consider the same three types of omic data (gene expression,
methylation, miRNA expression), but use different numbers of data sets (34 in \ngu vs. 10 in \rapp). 
Neither \ngu nor \rapp explicitly comment on the number of data sets and the selected cancer types, although 34 seems to be close to the maximum number of available data sets for the three considered types of omic data at the time of publication. Moreover, neither \ngu nor \rapp discuss their choice of omic data types, which seems to be a general issue in papers proposing new cancer subtyping methods, as criticized by \cite{Duan2021}.

Although the ten cancer types included by \rapp are also considered in \ngu, the corresponding data sets have different numbers of patients and omic variables. This is mainly caused by the different pre-processing steps applied by \ngu and \rapp (see supplement Section \ref{appendix:omics:prepro} for details). In addition, the two papers probably also use different data set versions (it was not possible to identify the data version used by \ngu).

Note that \ngu also consider two breast cancer data sets that do not originate from TCGA and exhibit different omic types. However, we exclude them from our experiment since some evaluation criteria of \rapp require six clinical variables (see below), which are either not available or cannot be clearly identified for these two data sets. Moreover, we do not include the partial data sets (i.e., data sets where some patients do not have any measurements for one or more omic data types) used in \rapp to demonstrate NEMO's ability to analyze this kind of data since PINSPlus assumes complete data and would require potentially suboptimal solutions such as imputation. \\

\noindent\textbf{Competing methods}\space\space
\rapp and \ngu use different numbers and types of competing methods to assess the relative performance of their proposed methods. While \rapp use nine competing methods, \ngu only consider three methods. The only method that is included in both papers is Similarity Network Fusion (SNF; \citealp{wang2014similarity}). The difference in the number of competing methods is not surprising given that the performance evaluation of NEMO is, in contrast to PINSPlus, based on a benchmark study with a focus on method comparison itself \citep{Rappoport2018}. Such studies typically aim to compare as many methods as possible to generate comprehensive guidelines for method users. Interestingly, \rapp include PINS, the predecessor method of PINSPlus, as a competing method. PINSPlus itself is not included since it did not exist yet when \cite{Rappoport2018} conducted their benchmark study. Concerning the choice of competing methods, \cite{Rappoport2018} report that they aim to represent diverse multi-omic clustering approaches, and that within each approach, they choose widely used methods with available software and clear usage guidelines. \ngu refer to the selected competing methods as established subtyping methods.

Regarding the parameter selection of the competing methods, NEMO's authors state in \cite{Rappoport2018} that they choose the method parameters following the guidelines given by the authors of the respective method (which involves performing a parameter search if suggested) and construct parameter selection methods by themselves if there are no available guidelines. \ngu do not have a comparable statement except for the number of clusters for the method Consensus Clustering (\citealp{monti2003consensus}), which, as stated in \cite{Nguyen2017}, is determined as suggested by \cite{monti2003consensus}.  
For SNF, the only method that is considered as competing method for both \pinsplus and \nemo, \ngu and  \rapp both normalize the omic variables to have a mean of 0 and a standard deviation of 1 (which, as stated in Section \ref{section:experiment_term}, we consider as a method parameter since it is not applied for all methods in both papers). However, they choose different values for the number of neighbors (20 vs.\ number of samples/10), the number of iterations (10 vs.\ 30), the number of clusters (estimate according to eigen-gaps vs.\ rotation cost), and the maximum number of considered clusters (5 vs.\ 15). See Table \ref{tab:design_multiomics_supp} for the method-specific pre-processing steps as well as \ngu and \rapp for all other parameters of the remaining methods. 

Note that we have to exclude two competing methods (rMKL-LPP and MultiNMF) considered by \rapp from the experiment since we are not able to run them (see supplement Section~\ref{appendix:omics:fail} for details).\\

\noindent\textbf{Evaluation criteria}\space\space 
With regard to the evaluation criteria, \ngu focus on the methods’ ability to identify clusters with significant survival differences using the logrank test. Note that in this context, the logrank test is equivalent to performing a Cox regression (which is the term used by \ngu), but we will refer to it as logrank test since this seems to be the more commonly used term in cancer subtyping methodology.  \cite{Nguyen2017} note that the same logrank test was also used by the authors proposing SNF \citep{wang2014similarity}, which can be seen as a justification for their choice. For each method, \ngu highlight the data sets with significant (i.e., $p < 0.05$),  and most significant (i.e., the smallest significant $p$-value across all methods) $p$-values by color.

In \rapp, the assessment of significant survival differences is also based on the logrank test. In addition, the authors assess \lq\lq clinical enrichment'' by testing the association between the identified clusters and six clinical variables (gender, progression of the tumor, cancer in lymph nodes, metastases, total progression, age at initial diagnosis), although not all variables are available in each clinical data set. \rapp employ the $\rchi^2$ test for independence for discrete and the Kruskal-Wallis test for continuous variables, and additionally adjust for multiple testing using the Bonferroni correction.
In contrast to \ngu,  \rapp estimate the logrank $p$-values using a permutation procedure, arguing that in the cancer subtyping context, the $\rchi^2$ distribution assumed for the logrank test statistic is often an inaccurate approximation and leads to increased type 1 errors (see supplement Section \ref{appendix:omics:permutation} for details). The same applies to the  $\rchi^2$ test for independence and the Kruskal-Wallis test. For each method, \rapp aggregate the individual results of each data set by reporting the number of data sets with significant logrank $p$-values, the number of data sets with at least one enriched clinical variable, the mean $-\mathrm{log}_{10}$ logrank $p$-value, and the mean number of enriched clinical variables per data set. \rapp thus consider four evaluation criteria regarding survival and clinical enrichment. Note that one of these criteria (number of data sets with significant logrank $p$-values) is very similar to the criterion used by \ngu (number of data sets with [most] significant logrank $p$-values), the only difference being the estimation of the $p$-value (approximation-based vs. permutation-based) and the inclusion of the number of data sets with the most significant $p$-values as second order ranking criterion in \ngu.

In addition to analyzing survival differences and clinical enrichment, \rapp also report the number of clusters and the runtime of each method. However, we do not consider these criteria in our experiment since the number of clusters has no clear optimal value and runtime is not comparable due to different computational resources.

 \subsection{Challenges when conducting the experiment} \label{section:challenge_omics}
\noindent\textbf{Reproducibility}\space\space  The results presented in \ngu are fully reproducible, except for one $p$-value of \iclusterplus. In contrast, the results presented in \rapp cannot be exactly reproduced. Besides the two methods that cannot be run at all, the performance results of the remaining methods are slightly different compared to the original paper, especially for the clinical enrichment criteria (the difference between original and reproduced results with regard to NEMO's performance is reported in Section~\ref{section:results_omics}). Interestingly, 76 of the 80 clustering solutions (8 methods $\times$ 10 data sets) are equal to the clustering solutions provided by \cite{Rappoport2018}, with two of the remaining four solutions only differing in one and three individuals, respectively. This means that the reproducibility problems (also observed for some of the 76 settings yielding identical clustering solutions) might be caused by the permutation tests. Moreover, the provided code is probably not the exact code used by \rapp, as indicated by the fact that \rapp refer to \cite{Rappoport2018} for the code to reproduce the results, but also mention that the implementations for MCCA, LRAcluster and $k$-means were slightly changed compared to \cite{Rappoport2018}.

When reproducing the results, we do not modify the code provided by the authors in a way that would change the results. However, we have to set a different number of cores in some settings and use a different \texttt{R} version for running the permutation tests by \rapp due to different computational resources (see our code documentation for details), which may contribute to the reproducibility issues. \\

\noindent\textbf{Crossing the designs}\space\space 
Evaluating the performance of \pinsplus and \nemo using each other's data sets, competing methods, and evaluation criteria poses a number of challenges. The most important one being the choice of parameters both for the two methods of interest, \pinsplus and \nemo, and the competing methods. Whenever a method is applied to a new (set of) data set(s), the method user needs to carefully select its parameters or a corresponding parameter selection method, which of course also applies to our experiment. Since both \ngu and \rapp use the same three types of omic data from the same source (TCGA), we set the parameters of \pinsplus and \nemo as in their respective original paper, which corresponds to their default parameter setting. Note that we also do not change the range of possible values for the number of clusters, a parameter that can be specified for both methods and is set to $[2,5]$ for PINSPlus and to $[2,15]$ for NEMO. We also attempt to use the same parameters for the competing methods when applying them to the new data sets. However, for two competing methods of \ngu (\iclusterplus and Consensus Clustering), the optimal number of clusters has to be selected by the user according to plots generated by the method when run on a specific data set. When applying these two methods on the data sets by \rapp, we thus have to manually choose the optimal number of clusters for every data set, and although we try to imitate the decisions of \ngu on their data sets, a clear determination is not always possible (an issue that is also noted by \citealp{Duan2021}). Moreover, some refinements regarding the method-specific pre-processing steps are necessary for two competing methods of \rapp (see Section \ref{appendix:omics:prepro} in the supplement).

In addition to the choice of method parameters, some challenges arise when applying the evaluation criteria  by \rapp on the data sets by \ngu. Specifically, the logrank permutation test by \rapp does not converge for some methods on two data sets by \ngu, resulting in a $p$-value of 0 in 15 method-data-combinations. In these cases, we use the approximation-based logrank $p$-values. Moreover, clustering solutions resulting from the data set UCS (\ngu) are not tested for clinical enrichment (\rapp) since it only includes one of the six clinical variables (``gender'') with only one value (“female”).

\subsection{Results}\label{section:results_omics} 

\begin{figure}
    \centering
  \includegraphics[width=0.9\textwidth]{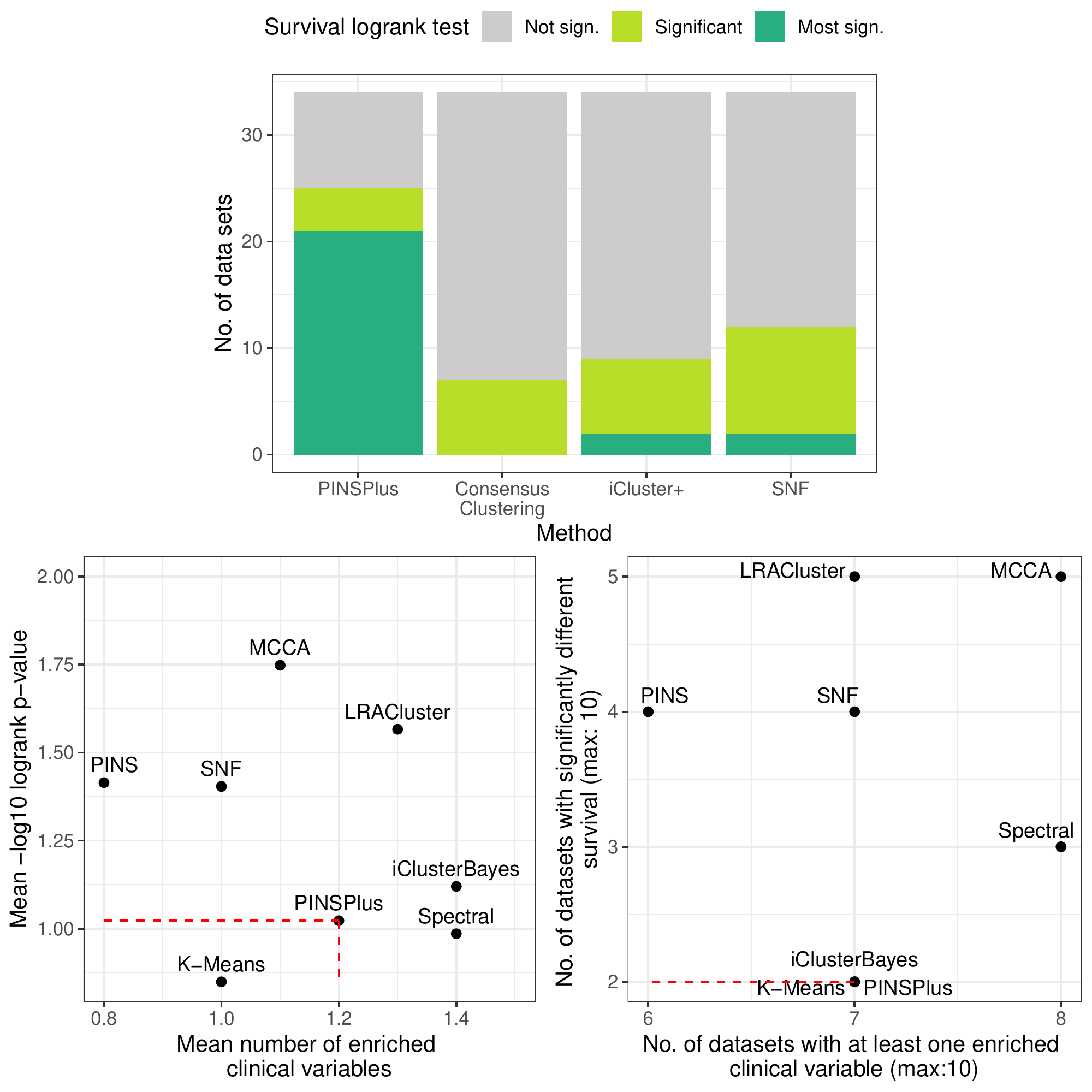}
       \caption{Performance of PINSPlus based on the original study design by \cite{Nguyen2019} (upper panel) and the study design by \cite{Rappoport2019} (lower panel).}
    \label{fig:pinsplus}
\end{figure}

\begin{figure}
    \centering
  \includegraphics[width=0.9\textwidth]{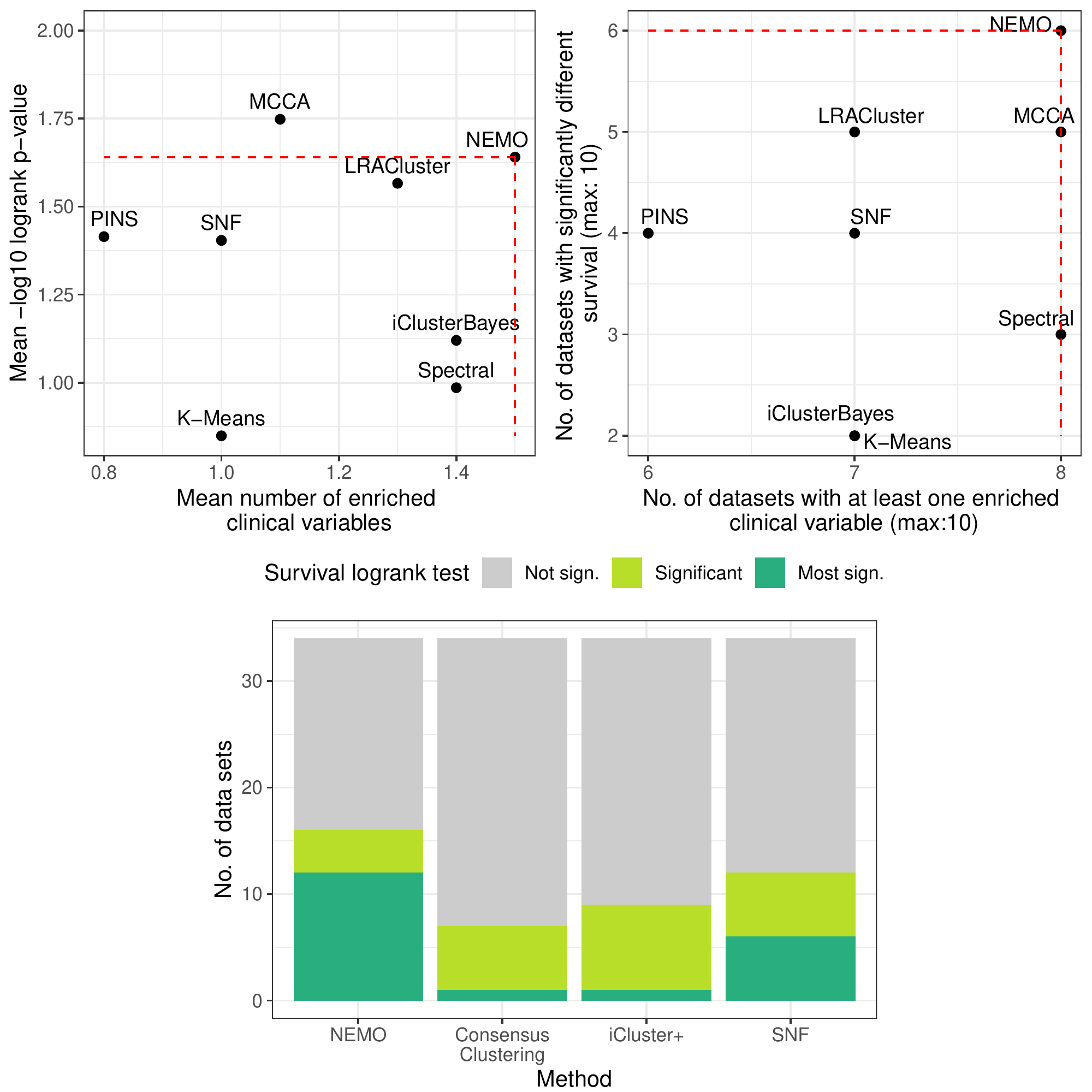}
       \caption{Performance of NEMO based on the original study design by \cite{Rappoport2019} (upper panel) and the study design by \cite{Nguyen2019} (lower panel).}
    \label{fig:nemo}
\end{figure}

\noindent\textbf{Performance based on the original study design}\space\space  The upper panels of Figure~\ref{fig:pinsplus} and \ref{fig:nemo} show the reproduced performance results of \pinsplus and \nemo based on their original study design. Note that the representation in thes figures slightly differs from the original papers to achieve a comprehensive and yet clear summary of the results. The most important difference is that the papers also report the individual performance results for each data set (we provide the individual performance results in Table~\ref{appendix:tab:pinsplus} and \ref{appendix:tab:nemo} in the supplement). 

When evaluated based on its original design, PINSPlus seems to be clearly superior to the three competing methods. It has the most significant  $p$-values ($p <0.05$) regarding survival, with 21 of the 25 significant $p$-values being the smallest across all methods. 
NEMO also shows good performance in its original study design, although its performance is not as clearly superior to the competing methods as the performance of PINSPlus. It achieves the highest numbers of data sets with significantly different survival and at least one enriched clinical variable (although there are two competing methods that achieve the same number of data sets with clinical enrichment). Moreover, none of the competing methods achieves both a higher mean $-\mathrm{log}_{10}$ logrank $p$-value and a higher mean number of enriched clinical variables. Only MCCA obtains a higher mean $-\mathrm{log}_{10}$ logrank $p$-value than \nemo but has a lower mean number of enriched clinical variables. Note that despite the reproducibility issues, both the absolute (i.e., the values of the four evaluation criteria considered by \rapp) and the relative performance of NEMO (i.e., when comparing these values to the competing methods) correspond to the results shown in the original paper. The only difference affecting the relative performance of \nemo is that in the original paper, one of the two methods that could not be reproduced (rMKL-LPP), achieves a higher mean number of enriched clinical variables than NEMO but a lower mean $-\mathrm{log}_{10}$ logrank $p$-value. \\

\noindent\textbf{Performance based on the crossed design}\space\space 
The performance results of \pinsplus and \nemo based on each others' study design (i.e., data sets, competing methods, and evaluation criteria) are presented in the lower panels of Figure \ref{fig:pinsplus} and \ref{fig:nemo}. 
In the study design of \rapp, PINSPlus does not outperform the competing methods. It is only the fourth and sixth best method with regard to the mean number of enriched clinical variables and mean $-\mathrm{log}_{10}$ logrank $p$-value, respectively. It belongs to the three worst methods with regard to the number of data sets with significantly different survival and only outperforms PINS, its predecessor method, with regard to the number of data sets with at least one enriched clinical variable. In contrast, NEMO still outperforms the competing methods in the design by \ngu, although its superiority is not as pronounced as for PINSPlus in the same design (PINSPlus achieves 25 significant $p$-values while NEMO only achieves 16 for the same 34 data sets). 

\begin{figure}
    \centering
    \includegraphics[width=1\textwidth]{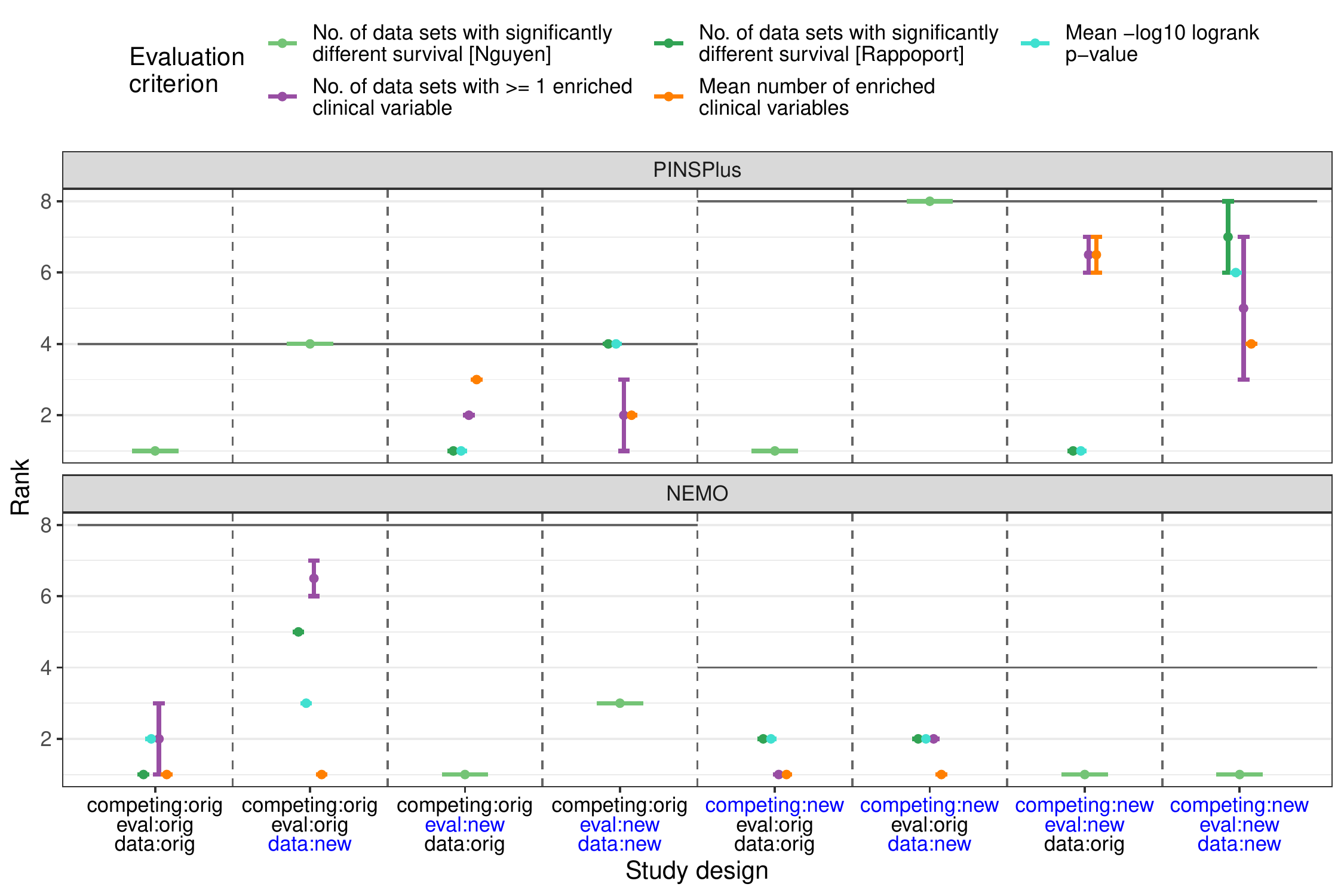}
    \caption{Performance ranks of NEMO and PINSPlus based on data sets, competing methods, and evaluation criteria that either correspond to the original (\pinsplus: \citealp{Nguyen2019}; \nemo: \citealp{Rappoport2019}) or crossed design (\pinsplus: \citealp{Rappoport2019}; \nemo: \citealp{Nguyen2019}). If more than one method achieves the same value for a certain criterion, the minimum, maximum and average rank are reported. The black lines correspond to the number of compared methods, i.e., the highest possible rank.}
    \label{fig:multiomics_ranks}
\end{figure}
We also analyze the performance of \pinsplus and \nemo when data sets, competing methods, and evaluation criteria are varied individually. Figure \ref{fig:multiomics_ranks} shows the resulting ranks of PINSPlus and NEMO for all eight combinations of the three components, where each component can either be set to the original or the crossed version ($2^3 = 8$). For each criterion (one by \ngu and four by \rapp), a rank of 1 corresponds to the best method. If more than one method achieves the same value for a certain criterion, the minimum, maximum, and average rank are reported. As can be seen from Figure \ref{fig:multiomics_ranks}, the ranks of PINSPlus and NEMO generally vary for each combination of data sets, competing methods, and evaluation criterion.
Apart from its original design, PINSPlus achieves rank 1 for the evaluation criteria related to survival (i.e., number of [most] significant $p$-values and mean $-\mathrm{log}_{10}$ logrank $p$-value) in all combinations where the data sets by \ngu are used. However, PINSPlus belongs to the worst performing methods according to survival when applied to the data sets by \rapp. As mentioned in Section~\ref{section:design_omics}, the ten data sets corresponding to different cancer types that are used by \rapp are also included in \ngu. Interestingly, PINSPlus achieves a significant $p$-value for nine of these ten data sets in \ngu, indicating that the difference in performance for these data sets is mainly due to the different pre-processing steps. With regard to the clinical evaluation criteria, \pinsplus seems to have average performance, neither clearly performing better nor worse than the other methods. 

In comparison to PINSPlus, the ranks of NEMO are more robust across the different study designs. For six of eight study designs, it achieves rank 1 or 2 for all evaluation criteria (if the minimum or average rank is considered). The only study design where NEMO’s performance is considerably worse for two evaluation criteria is the design where only the data sets are taken from \ngu while evaluation criteria and competing methods correspond to the original paper. Moreover, it can be noted that while the slightly different calculation of the number of data sets with significant logrank p-values in \ngu and \rapp does not have an impact on the ranks of PINSPlus, NEMO tends to achieve better ranks for the version of \ngu. For example, it achieves rank 1 instead of 2, for settings where data and competing methods are by \ngu. A comparison of approximation-based and permutation-based $p$-values for all methods and data sets can be found in the supplement (Figure \ref{fig:appendix:pvalue}), showing that the approximation-based $p$-values are indeed generally smaller. The supplement also provides a comparison of the two different parameter settings of SNF that are specified by \ngu and \rapp (Figure \ref{fig:appendix:snf}), which reveals a considerable but non-systematic performance difference between the two implementations.

\section{Data analysis task II: Differential gene expression analysis}\label{section:de}
The second data task we consider in our experiment is differential gene expression analysis, which aims at identifying genes that show differences in their expression levels between two or more conditions \citep{Soneson2013}. Of the many methods that have been proposed for this task \citep{Seyednasrollah2013}, the more recent ones usually expect RNA-Seq data as input, which means that gene expression is measured as non-negative counts \citep{Rigaill2018}. The two methods for differential expression analysis included in the experiment are \sfmeb and \mbcdeg, which have recently been proposed by \cite{Zhou2021} and \cite{Osabe2021} and require RNA-Seq data as input. As stated in Section~\ref{section:experiment}, these papers are selected because they make the code to reproduce the results openly available (information on where the code can be found is reported in our code documentation). We will abbreviate them by \zhou and \osabe in the following.

\subsection{Study design in the original papers}\label{section:design_de}
In this section, we review the data sets, competing methods, and evaluation criteria that are used to asses the performance of \sfmeb and \mbcdeg in their respective original paper and that meet the inclusion criteria of our experiment (see Table~\ref{tab:design_de} for an overview). We also report the  justifications of the design choices provided by the authors. Since \zhou and \osabe primarily use simulated data to evaluate their methods, we do not further consider their real data analyses. \\

\begin{table}[ht!]\footnotesize 
\caption{Overview of the study design components  used for performance assessment of \sfmeb and \mbcdeg. Included are only components  (i) for which the corresponding results are presented as figures or tables in the main paper (i.e., not in the supplement), (ii) that compare the method of interest to other competing methods, and (iii) that correspond to the performance evaluation based on simulated data.  In addition, some components are not included in the experiment, which are indicated by asterisks (*). Competing methods and evaluation criteria for data sets not included in the experiment are not shown. In case of design-implementation-gaps, the  table  refers to the code for reproducing the results. }
\label{tab:design_de}
    \begin{tabularx}{\textwidth}{|X|L{1.3cm}|L{1.18cm}|L{1.18cm}|L{1.18cm}|L{1.18cm}|L{1.35cm}|L{1.1cm}|L{1.1cm}|}
 \hline  
 \textbf{\raggedright Study design component} & 
 &
 \multicolumn{5}{|l|}{\thead[l]{\textbf{SFMEB} \\ \citep{Zhou2021}}}&
  \multicolumn{2}{|l|}{ \thead[l]{\textbf{MBCdeg} \\ \citep{Osabe2021}}}\\

\hline
\multirow{13}{4em}{\textbf{Data sets}}& 
No. of conditions &
\multicolumn{5}{|l|}{2 conditions} &
\multicolumn{2}{|l|}{ \thead[l]{2 conditions, \\ *3 conditions}} \\
\cline{2-9}
& 
No. of settings &
\multicolumn{5}{|l|}{15} & 
\multicolumn{2}{|l|}{24} \\
\cline{2-9}
& Setting names
&
Study 1 & 
Study 2 &
Study 3 & 
Study 4 &
Study 5 & 
Fig. 1 in \osabe&
Fig. 3 in \osabe \\
\cline{2-9}
&
Distribution & 
\multicolumn{4}{|l|}{Poisson} &
Negative binomial &
\multicolumn{2}{|l|}{Negative binomial} \\
\cline{2-9}
&
Code based on &
\multicolumn{4}{|l|}{Robinson and Oshlack (2010)} &
\raggedright\arraybackslash \texttt{R} package \texttt{compcodeR} & 
\multicolumn{2}{|l|}{\texttt{R} package \texttt{TCC}}\\ 
\cline{2-9}
&
No. of repetitions &
\multicolumn{5}{|c|}{20} &
100 & 50\\
\cline{2-9}
&
Samples per group &
\multicolumn{2}{|c|}{1} &
\multicolumn{3}{|c|}{\{2,5,8\}} &
\multicolumn{2}{|c|}{3} \\
\cline{2-9}
& 
Total no. of genes &
16500 &
28800 &
16800 &
29800 &
15000 &
10000 &
10000 \\

\cline{2-9}
&
UE genes  &
(1000,500) &
\centering (1000,500), (800,1500) &
(1000,800) &
\centering (1000,800), (2000,1000) &
- &
- &
- \\
\cline{2-9}
& 
Prop. of DE genes (excl. UE) & 
\{0.3, 0.5, 0.7\} &
0.6 $+$ \{0.1, 0.3, 0.5\} & 
0.6 &
0.6 $+$ 0.4 &
0.6 &
\{0.25, 0.05\} &
\{0.45,0.55, 0.65,0.75\} \\
\cline{2-9}
& 
Log$_2$ fold-change &
2 &
2 + 3 &
2 &
2 + 3 &
$\geq$ 3 & 
2 &
2\\
\cline{2-9}
& 
Prop. of up-regul. genes & 
0.9 &
0.9+ 0.1 &
0.6 &
0.9+ 0.1 &
1 &
\{0.5, 0.7, 0.9, 1\} &
\{0.5, 0.7, 0.9, 1\} \\
\cline{2-9}
& 
Heterog. data & 
no &
yes &
no &
yes &
no &
\multicolumn{2}{|c|}{no} \\
\cline{2-9}
&
Pre-filtering &
\multicolumn{4}{|l|}{\thead[l]{Mean count $\leq$ 2}}& 
\multicolumn{1}{|l|}{\thead[l]{Total\\ count 
$<$ 1}} & 
\multicolumn{2}{|l|}{-}\\
\hline
\multirow{3}{4em}{\textbf{Competing methods}} & 
Type and number &
\multicolumn{5}{|l|}{5: edgeR, DESeq/DESeq2, HTN, Library Size, NOISeq} &
\multicolumn{2}{|l|}{3: edgeR, DESeq2, TCC} \\ 
\cline{2-9}
 & 
Standard edgeR&
\multicolumn{2}{|l|}{No (binomial test)} &
\multicolumn{2}{|l|}{No (based on poisson)} &
\multicolumn{1}{|l|}{Yes} &
\multicolumn{2}{|l|}{Yes} \\ 
\cline{2-9}
& 
Other method parameters &
\multicolumn{5}{|l|}{See original paper} &
\multicolumn{2}{|l|}{See original paper} \\ 

\hline
\multirow{2}{4em}{\textbf{Evaluation criteria}} & 
Metric & 
\multicolumn{4}{|l|}{AUC} &
AUC (smoothed ROC)
& 
\multicolumn{2}{|l|}{AUC} \\
\cline{2-9}
& 
Aggregation &
\multicolumn{5}{|l|}{Boxplots} & 
\multicolumn{2}{|l|}{Boxplots} \\
\hline
\end{tabularx}
\end{table}

\noindent\textbf{Data}\space\space 
Both \zhou and \osabe generate simulated count data representing RNA-seq read counts of $p$ genes in $2 \times n_{obs}$ samples from two groups. 
\osabe also simulate count data from three groups, but we exclude these settings from the experiment because \sfmeb does not seem to be intended for this type of data (all evaluations in the original paper by \zhou are based on two-group data). The simulation framework of \zhou and \osabe is based on different code implementations (code by \citealp{Robinson2010} and {\texttt{compcodeR}} \texttt{R} package, \citealp{compcodeR} vs. {\texttt{TCC}} \texttt{R} package,  \citealp{TCCpackage}) 
as well as different distributions to generate the count data (Poisson and negative binomial distribution  vs. only negative binomial distribution). Moreover, the two papers choose  different numbers of simulation repetitions (20 vs. \{50,100\}), different sample sizes per group (\{1,2,5,8\} vs. 3), and different numbers of genes (\{15000,\dots, 29800\} vs. 10000).

 The simulations also differ with respect to the characteristics of the differentially expressed (DE) genes. In contrast to \osabe, the DE genes in \zhou include uniquely expressed (UE) genes $(u_1,u_2)$ that have zero counts in group 1 or 2, respectively. Moreover, \zhou and \osabe consider different proportions of DE genes (\{0.3,\dots,0.7\} excluding UE genes vs. \{0.05,\dots,0.75\}),  different $\mathrm{log}_2$ fold-changes between the groups (i.e., the true $\mathrm{log}_2$ ratio of expression change; $\geq$ 2 vs. 2), and different proportions of up-regulated genes (i.e., genes having higher expression) in group 1 (\{0.6,\dots,1\} vs. \{0.5,\dots,1\}). 

In contrast to \osabe, \zhou apply pre-filtering of the genes (e.g.,  filtering of genes with mean count $\leq$ 2) for all methods, although some of their considered methods additionally filter genes internally. Moreover, in some settings, \zhou consider heterogeneous data composed of two data sets with different simulation parameters ($\mathrm{log}_2$ fold-change, number of genes, etc.). 
In the results included in the experiment, \osabe only vary the proportion of DE genes and the proportion of up-regulated genes, but in a fully factorial manner which results in $6\times4=24$ simulation settings. However, it should be noted that \osabe also vary other parameters (e.g., the $\mathrm{log}_2$ fold-change) in settings not considered in our experiment since they did not meet the inclusion criteria (e.g., because the corresponding figures are shown in the supplement). In the simulation settings by \zhou that are included in our experiment (15 settings in total), more parameters are varied, but not in a fully factorial manner. More specifically, the 15 included settings originate from five ``studies'' (each consisting of 3 settings) with different simulation parameters. Within each study, one simulation parameter is varied (see Table \ref{tab:design_de}). 

Understandably, neither \zhou nor \osabe provide a justification for every single simulation parameter, but often refer to similar parameter values observed in real data. Regarding the choice of the number of simulation repetitions, however, neither of the two papers provides a justification. As criticized by \cite{morris2019using}, this seems to be a general issue in papers presenting simulation studies.\\

\noindent\textbf{Competing methods} \space\space 
\zhou compare \sfmeb with five competing methods they consider as widely used. Two of these methods are referred to as edgeR and DESeq (\citealp{Robinson2009}; \citealp{Anders2010}; see below for more details), which closely corresponds to the methods selected by \osabe (edgeR and DESeq2; \citealp{Love2014}). In addition to these two methods, \osabe also consider the less well-known method TCC \citep{TCCpackage}, arguing that it is not sufficient to compare a newly proposed method to the most commonly used methods (edgeR and DESeq2) as those might not be the ones best suited for the analysis. Moreover, they see TCC as the main alternative to their proposed method since the normalization algorithm used by TCC corresponds to the normalization algorithm used by one version of \mbcdeg. 

Interestingly, \zhou and \osabe use different implementations of edgeR. While the implementation by \osabe corresponds to one of the edgeR standard workflows, \zhou use three different implementation of edgeR across their simulation settings of which only one would be typically considered as edgeR (but still with different parameters than \osabe), while the other two are only edgeR-like.
One reason for this choice is that some simulation settings in \zhou do not have biological replicates (i.e., $n_{obs}=1$ in each group), for which the standard edgeR implementation yields an error (see supplement Section \ref{appendix:de:edger} for details). Regarding the implementation of DESeq/DESeq2, \zhou actually use both DESeq and DESeq2, although they generally refer to the method as DESeq, the predecessor method of DESeq2. This might be explained by the fact that, similar to edgeR, DESeq2 is not intended for settings without biological replicates and thus yields an error, which is why \zhou use DESeq in these settings. Note that it has been shown that DESeq and DESeq2 perform differently \citep{Love2014}. Both \zhou and \osabe use the same parameters for DESeq2.
For the parameters of the remaining methods see \zhou and \osabe as well as the referenced code.\\

\noindent\textbf{Evaluation criteria}\space\space 
Both \zhou and \osabe assess the methods' ability to correctly identify DE genes using the area under the receiver operating characteristic curve (AUC). They both justify this decision with the fact that the AUC does not require the choice of a threshold value as other popular measures. The AUC takes values from 0 to 1, where 1 corresponds to perfect discrimination of DE and non-DE (i.e., non-differentially expressed) genes, and 0.5 corresponds to random assignment. However, due to an unfortunate default option in the \texttt{R} package used by \zhou to calculate the AUC, the resulting AUC values are 1 minus the correct AUC for some repetitions (see supplement Section \ref{appendix:de:auc} for details). Apart from the different \texttt{R} packages used to calculate the AUC, \zhou also employ a \textit{smoothed} ROC curve to estimate the AUC in some of their simulation settings (study 5), which can lead to slightly different results compared the non-smoothed ROC curve. 
Regarding the aggregation of AUC values across the simulation repetitions, both \zhou and \osabe use boxplots.

 \subsection{Challenges when conducting the experiment} \label{section:challenge_de}
\noindent\textbf{Reproducibility}\space\space 
When reproducing the results presented in \osabe and \zhou, we do not modify the original code in a way that would change the results, with one exception: We change the number of simulation repetitions from 10 to 20 (i.e., the number reported in the paper) in the code provided by \zhou since the results using 20 repetitions are more similar to the results shown in \zhou (note that we make this change before crossing the designs). As stated in Section \ref{section:experiment}, we also use the same \texttt{R}  and  \texttt{R} package versions as in the original papers. However, \zhou do not provide this information, which is why we use the most recent package versions available when conducting the experiment (see our code documentation for the exact version information). The code by \zhou also does not include a random number seed, which we accordingly set but which is most likely different from the seed used by \zhou. Note that for reproducing the results of \zhou, we use their AUC implementation potentially yielding incorrect results, but additionally calculate the correct version. 

Based on these modifications, running the code of  \zhou and \osabe results in very similar but not exactly the same boxplots as shown in the original papers. More specifically, the relative performance of each method is the same in the original and reproduced versions, but some boxplots have e.g. different outliers.  For \zhou, this relatively high degree of reproducibility is noteworthy considering the fact that the provided code does not include a seed or version information. The only three settings that do not yield similar results are the settings from study 5 by \zhou (the differences between the original and reproduced results are described in Section \ref{section:results_de}). Apart from the aforementioned missing seed and version information, the different results in these settings could be due to the fact that the code might not have been provided in its final version. \\

\noindent\textbf{Crossing the designs}\space\space 
As already stated in the first example on cancer subtyping, conducting the cross-design experiment implies that all considered methods are applied to new data sets (new in the sense that these data sets have not been included in the original paper). It is thus necessary to carefully specify the method parameters of \sfmeb, \mbcdeg, and all competing methods. Although the simulation settings of \zhou and \osabe are less comparable than the real data sets of \ngu and \rapp in the cancer subtyping example, we nevertheless adopt the parameter values from the original papers because we consider the risk of running the methods with suboptimal parameter settings to be lower for the parameters used by \zhou and \osabe than for parameters selected by ourselves (especially because we select the parameters \textit{before} seeing the results to avoid the risk of favoring one of the methods, as stated in Section~\ref{section:experiment_design}).
However, both \zhou and \osabe consider more than one parameter value for some methods and \zhou even use different methods across the simulation settings (i.e., DESeq and DESeq2). For all methods evaluated in \zhou (i.e., \sfmeb and its competing methods), we adopt the parameters from study 5 since they are the most similar to the simulation settings considered in \osabe (i.e., non-heterogeneous data, generated using the binomial distribution, with replicates). In all simulation settings of \osabe included in our experiment, the authors evaluate two versions of \mbcdeg, which are denoted as \mbcdega and \mbcdegb and correspond to two different normalization options. Since \mbcdega and \mbcdegb are also implemented separately in the code, we include both versions in the experiment but decide to focus on \mbcdegb, which was observed to be slightly more stable and accurate in \osabe, before seeing any results. Although \osabe do not vary any other parameters of \mbcdeg or the competing methods, we note that the main parameter of \mbcdeg that is extensively discussed by \osabe might not be ideal for some simulation settings of \zhou. We thus conduct a sensitivity analysis using two different values for this parameter (see Section~\ref{appendix:de:sa} for details).

Since \osabe and \zhou use the same evaluation criterion (i.e., boxplots representing the AUC of each simulation repetition), we only re-evaluate the performance of \sfmeb and \mbcdeg on each other’s competing methods and data. When crossing the designs, we do not consider the AUC that is based on the smoothed ROC curve used by \zhou in some simulation settings. Of course, we also do not use the incorrectly calculated version of the AUC.

Note that not all design components of \zhou and \osabe are compatible. More specifically, the DESeq2 and edgeR implementation in \osabe results in an error when applied to the simulation settings without biological replicates by \zhou. As stated in Section~\ref{section:design_de}, this is because DESeq2 and edgeR are not intended for settings without biological replicates and \osabe do not use a (possibly non-ideal) workaround solution as done by \zhou.

\subsection{Results} \label{section:results_de}

\begin{figure}
    \centering
      \includegraphics[width=0.78\textwidth]{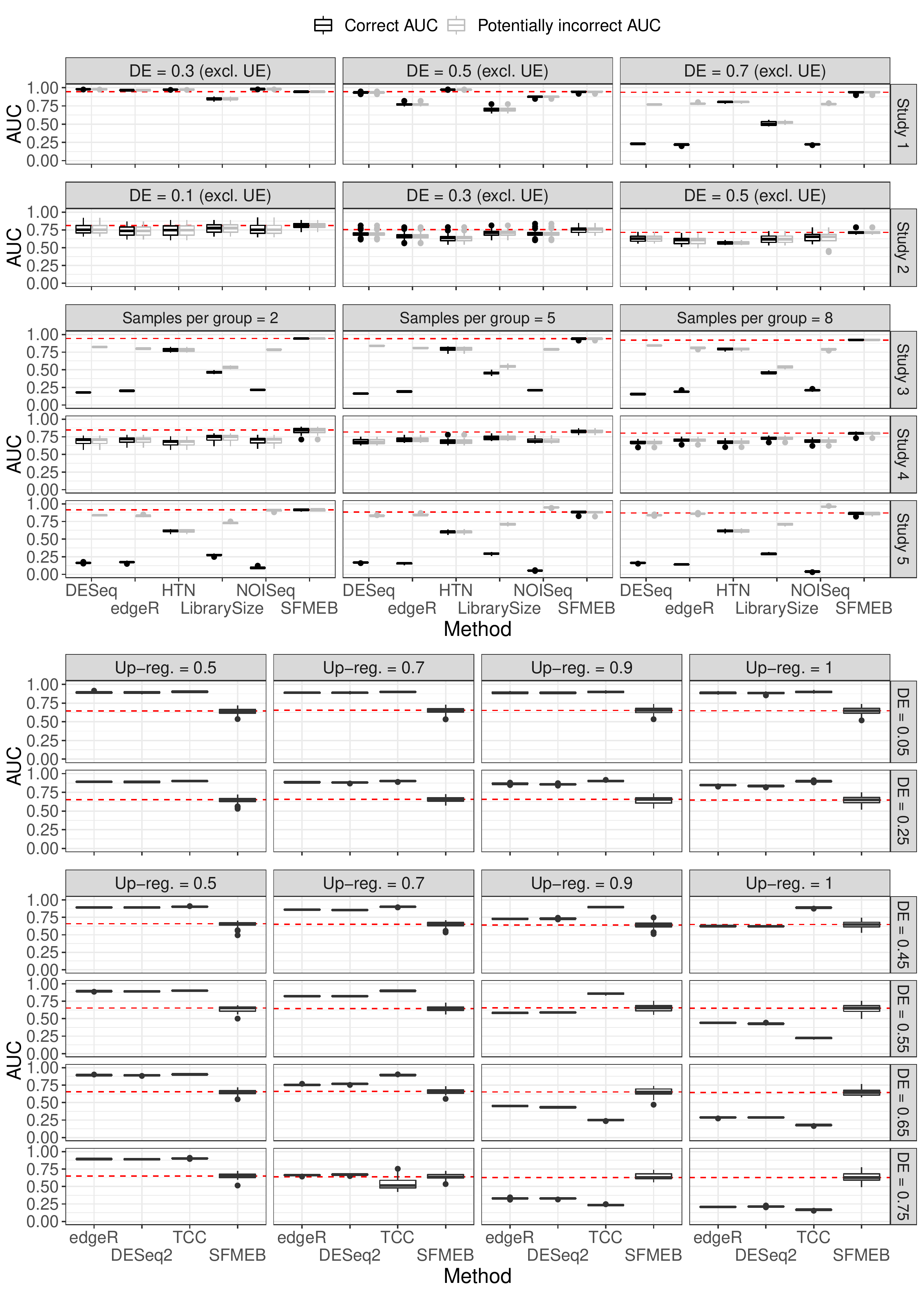}
         \caption{Performance of \sfmeb based on the original study design by \cite{Zhou2021} (upper panel) and the study design by \cite{Osabe2021} (lower panel). The boxplots correspond to $n_{sim}$ simulation repetitions, where $n_{sim} = 20$ for  \cite{Zhou2021} and $n_{sim} \in \{50,100\}$ for \cite{Osabe2021}. The red dashed line corresponds to the median AUC of \sfmeb across all simulation repetitions. In the original paper by \cite{Zhou2021}, the AUC has not been calculated as intended by the authors, which is why the correct AUC values are provided in addition to the reproduced and potentially incorrect AUC values.}
    \label{fig:sfmeb}
\end{figure}

\begin{figure}
    \centering
         \includegraphics[width=0.78\textwidth]{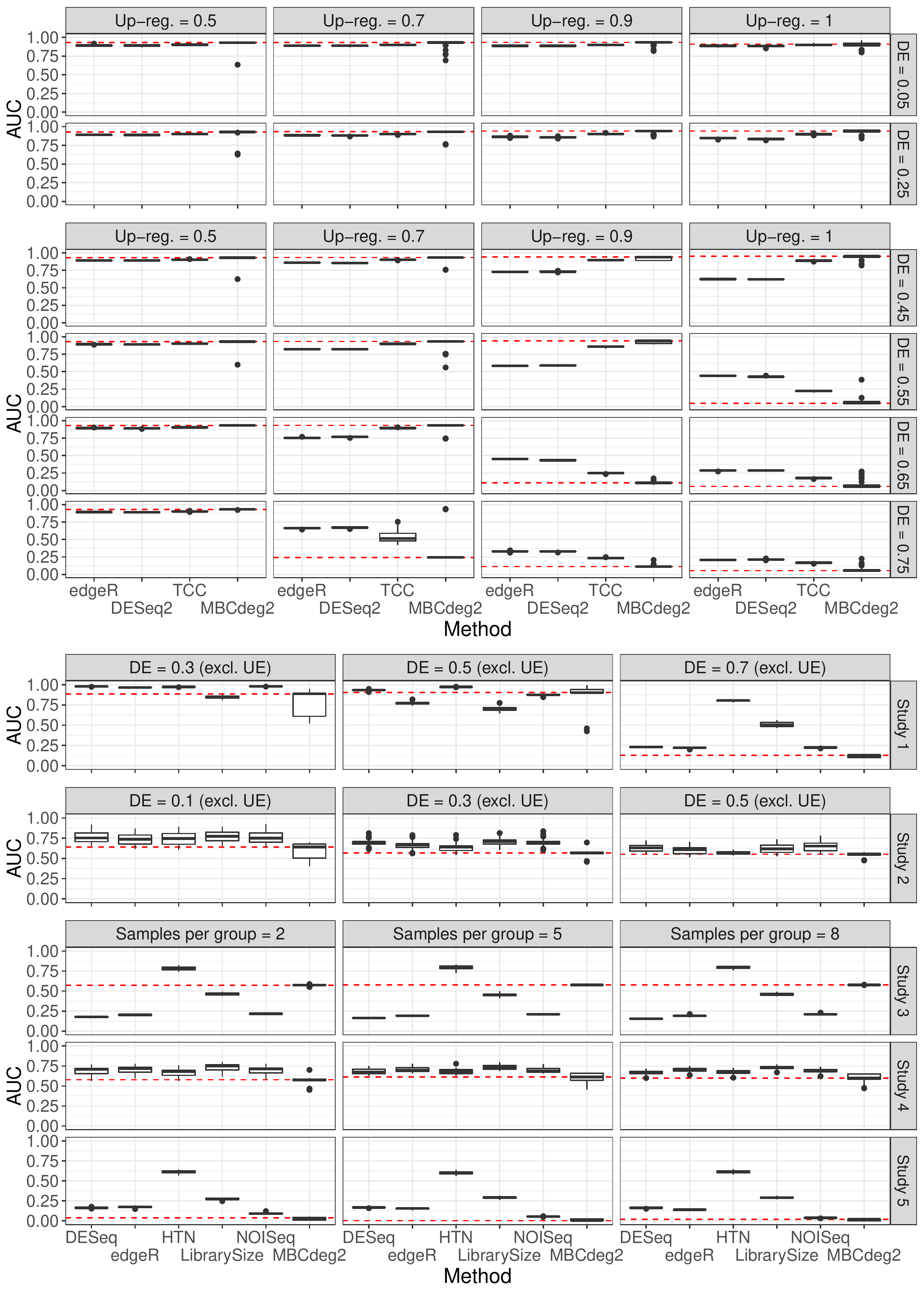}
         \caption{Performance of \mbcdegb based on the original study design by \cite{Osabe2021} (upper panel) and the study design by \cite{Zhou2021} (lower panel). The boxplots correspond to $n_{sim}$ simulation repetitions, where $n_{sim} \in \{50,100\}$ for \cite{Osabe2021} and $n_{sim} = 20$ for  \cite{Zhou2021}. The red dashed line corresponds to the median AUC of \mbcdegb across all simulation repetitions.}
     \label{fig:mbcdeg}
\end{figure}

\noindent\textbf{Performance based on the original study design}\space\space  
The upper panels of Figure~\ref{fig:sfmeb} and \ref{fig:mbcdeg} show the reproduced performance results of \sfmeb and \mbcdegb with an additional dashed line corresponding to the median AUC of the corresponding method of interest over all simulation repetitions. Note that the method labels are adopted from the original papers although the competing methods DESeq and edgeR in \zhou do not exactly correspond to the actual method in some simulation settings as discussed above. 

For \sfmeb, we show both the reproduced AUC values that are potentially biased towards higher values and the correct AUC values. As stated in the previous section, we only observe a noteworthy performance difference between the reproduced results and the results shown in \zhou for three simulation settings (i.e., study 5). In these settings, two competing methods consistently show better performance in the reproduced version, leading to \sfmeb being the second best instead of best performing method in two settings. However, these differences become irrelevant when looking at the correct AUC results. In fact, only the AUC values of the competing methods are in some settings affected by the incorrect AUC calculation, resulting in \sfmeb outperforming its competing methods more clearly than initially claimed by its authors. The performance results of \sfmeb based on the corrected AUC values are thus still consistent with the conclusion of \zhou that \sfmeb outperforms its competitors in most settings (achieving rank 1 according to median AUC in 13 out of 15 settings).

\mbcdegb also performs well in its original study design. As noted by \osabe, the method tends to achieve higher AUC values in the settings with a small ($\leq$ 0.45) proportion of DE genes. In some settings where the proportion of DE genes is $\geq 0.55$, however, the method seems to fail, often resulting in AUC values below 0.25 and not being able to outperform any of its competing methods (the same applies to  \mbcdega). \osabe discuss the occasional failure of \mbcdeg extensively and conclude that the identification of the non-DE gene cluster (which they state to be the key to the proposed framework) fails in these cases, which leads to an incorrect classification of DE and non-DE genes. 
However, \mbcdegb generally performs better than the competing method TCC in settings where TCC performs well (the same applies to \mbcdega). Given the fact that TCC could be expected to outperform other methods since the data sets are generated using the \texttt{TCC} \texttt{R} package and the normalization algorithm used by TCC was designed for settings with asymmetric (i.e., $\neq 0.5$) up-regulation as considered by \osabe, \osabe see this as the main contribution of their study. \\ %

\noindent\textbf{Performance based on the crossed design}\space\space 
The lower panels of Figure~\ref{fig:sfmeb} and \ref{fig:mbcdeg} display the performance results of \sfmeb and \mbcdegb based on each other's simulation data and competing methods. In the study design of \osabe, \sfmeb generally shows worse performance than in its original design, having lower median AUC values than all of its competitors in 17 out of 24 settings. However, in 5 out of the remaining 7 settings (the settings with a high proportion of DE genes that are mostly up-regulated in one group), \sfmeb clearly outperforms the competing methods. Interestingly, this difference in relative performance is mainly caused by the varying AUC values of the competing methods edgeR, DESeq2, and TCC. \sfmeb itself, on the other hand, shows very robust AUC values across all settings. However, with a median AUC of about 0.65 in each setting, \sfmeb's absolute performance is worse than in the original study, where the lowest median AUC of \sfmeb is 0.72.

Similar to \sfmeb, \mbcdegb generally performs worse compared to its original design. In 10 out of 15 settings, it is outperformed by all competing methods. However, it is the second best method in 4 of the remaining 5 settings (based on median AUC). In contrast to \sfmeb, the absolute performance varies more across the settings and only reaches a value comparable to the original design (excluding the settings where the method failed) in two settings. Similar to its original design, there are four settings where \mbcdegb shows extremely low AUC values, which again seems to be caused by the incorrect identification of the non-DE cluster (note that these are all settings where the proportion of DE genes is $\geq 0.6$, which is consistent with \osabe's observation in the original paper). As stated in Section~\ref{section:challenge_de}, we also conduct a sensitivity analysis where \mbcdegb's main parameter is set to a different value. However, this does not improve the AUC values (see Figure~\ref{fig:appendix:sa} in the supplement).

Figure \ref{fig:deanalysis_ranks} shows the resulting performance ranks of \sfmeb and \mbcdegb when data and competing methods are varied individually. For both \sfmeb and \mbcdegb, the data sets and competing methods can either be set to their original or the crossed version, which results in four ($= 2^2$) different study designs. Within each study design, the ranks are calculated separately for each simulation setting  based on the median AUC and are summarized as boxplots (i.e., each boxplot consists of 15 or 24 ranks, which corresponds to the total number of settings considered by \zhou and \osabe, respectively). All AUC values are calculated correctly. For both \sfmeb and \mbcdegb, the performance mainly depends on which simulated data sets are considered. In contrast, using different competing methods has no considerable impact on the distribution of ranks, except that the maximum possible rank reflecting the worst method varies according to the number of competing methods. This is also due to the partial overlap of competing methods in both designs. 

The results of \mbcdega based on the crossed design are very similar to the results of \mbcdegb and can be found in the supplement (Figure~\ref{fig:appendix:mbcdeg1}).

\begin{figure}
    \centering
    \includegraphics[width=0.8\textwidth]{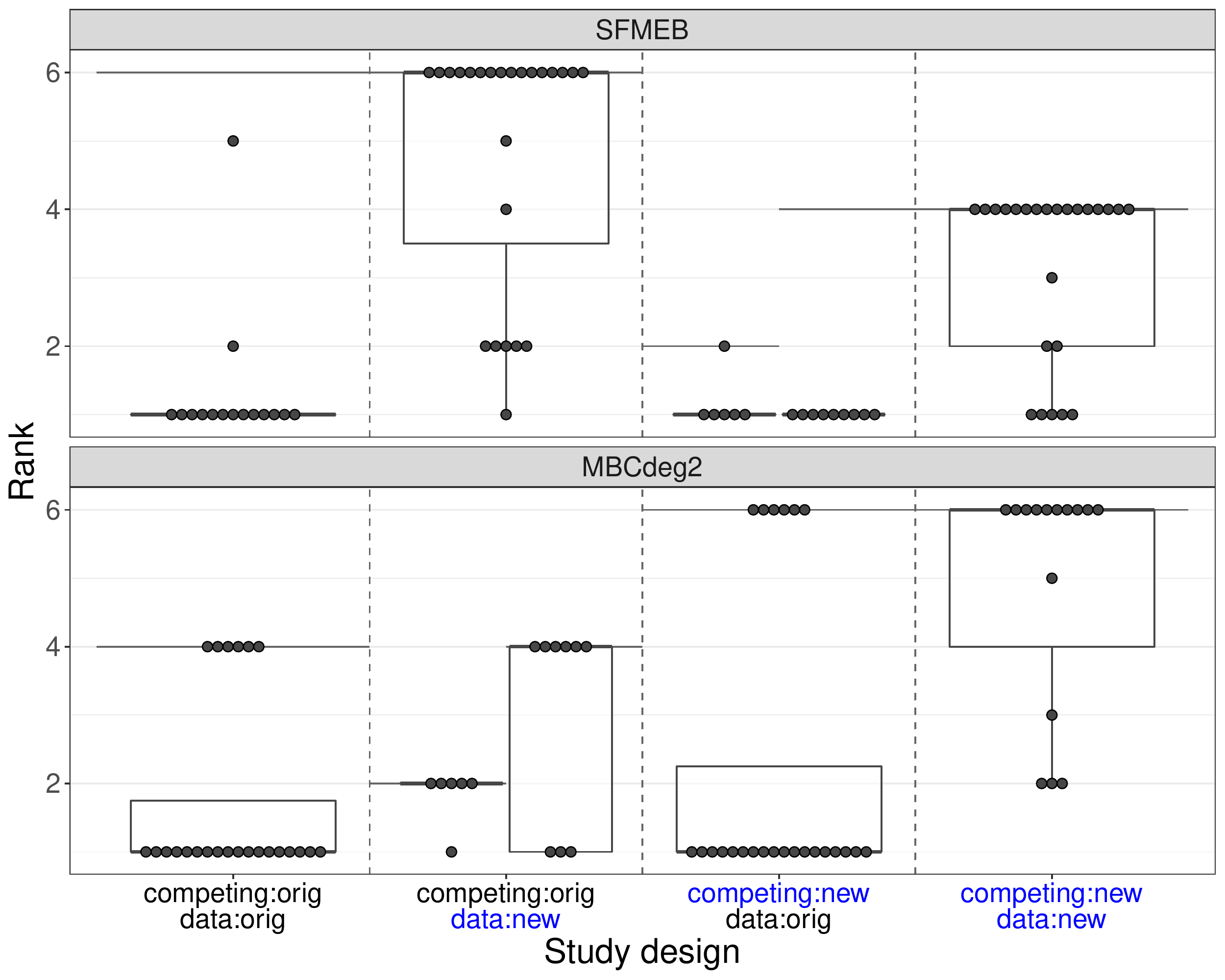}
  \caption{Performance ranks of \sfmeb and \mbcdegb based on data and competing methods that either correspond to the original (\sfmeb: \citealp{Zhou2021}; \mbcdegb: \citealp{Osabe2021}) or crossed design (\sfmeb: \citealp{Osabe2021}; \mbcdegb: \citealp{Zhou2021}). Each boxplot consists of 15 or 24 ranks (additionally represented as black points), which corresponds to the number of data settings considered by  \cite{Zhou2021} and \cite{Osabe2021}, respectively. The ranks of \sfmeb and \mbcdegb in each data  setting are calculated based on the median AUC value across all simulation repetitions. The black lines correspond to the number of compared methods, i.e., the highest possible rank. Note that for one combination of data and competing methods, not all competing methods can be applied to all settings, which is why the number of compared methods varies between 2 and 4.}
     \label{fig:deanalysis_ranks}
\end{figure}

\clearpage
\section{Discussion} \label{section:discussion}
\subsection{Summary of results and limitations} 
In this paper, we conducted a systematic experiment, which we refer to as ``cross-validation of methods'' and in which we re-evaluated methods based on the data sets, competing methods, and evaluation criteria of a paper proposing a method for the same data analysis task.
We considered two exemplary data analysis tasks, namely cancer subtyping using multi-omic data and differential gene expression analysis. For each analysis task, we selected two methods, \pinsplus \citep{Nguyen2019} and \nemo \citep{Rappoport2019} for cancer subtyping, and \sfmeb \citep{Zhou2021} and \mbcdeg \citep{Osabe2021} for differential expression analysis.

Although we did not conduct our cross-design validation experiment on a large scale, several interesting findings emerged. First, the difficulties in finding eligible papers showed that many papers are still being published without openly available code for reproduction. For the papers that were selected, running the provided code did not yield the exact same results as presented in the respective paper. The results of \pinsplus, however, were close to being fully reproducible with only one differing $p$-value in one of the competing methods. Nevertheless, the reproduced results of all four methods were consistent with the conclusion of the original papers that the respective method shows good performance. 

Second, the experiment concretely illustrated the researcher degrees of freedom regarding the performance assessment of a method. Notably, all four study designs seemed well though-out and the authors provided justifications in most cases. Interestingly, even for the design components that were similar in both papers, the exact implementation was often different. For example, SNF and edgeR were included as competing methods in both papers of the cancer subtyping and differential expression analysis task, respectively, but were run with different parameters.

Third, the experiment showed how differences in the study design can affect the performance of a method. Three out of the four considered methods (\pinsplus, \sfmeb, and \mbcdeg) performed  worse when assessed on the crossed study design. Only one method, NEMO, performed well when evaluated on the study design by PINSPlus and only showed slightly worse performance in some settings where data sets, competing methods, and evaluation criteria were varied individually. For both analysis tasks, using different data sets (real or simulated) had the largest impact on the performance results, which was particularly surprising for the real data sets of the cancer subtyping example where both papers used the same data type and source.

It is important to note that while the findings of our experiment might help to see the performance reported in the original papers from a different perspective, they cannot be seen as evidence of any of the four methods generally having good or bad performance. First, our experiment is limited in the sense that we did not include all study designs and corresponding results reported in the papers, which gives an incomplete picture regarding the study design of the papers and, importantly, the individual strengths and weaknesses of each method. This also includes qualitative evaluation criteria such as PINSPlus’ user-friendliness regarding the choice of the number of clusters (which was also noted by \citealp{Duan2021}), NEMO’s simplicity and support of partial data, the avoidance of potential error-prone data normalization when using \sfmeb, and the high interpretability of MBCdeg’s main parameter. Second, the method performances observed in the experiment clearly depend on (i) our own expertise regarding each method, and (ii) the respective new design we re-evaluated each method on (in principle, the $2\times2$ table in Section \ref{section:experiment} [Table \ref{tab:experiment}], could be extended to a $K\times K$ table by including further designs, and each of them would probably yield substantially different results). This can be seen as a limitation of our study.

\subsection{Mechanisms leading to an optimistic performance evaluation and possible solutions}
In our experiment, we observed that three out of four methods performed worse when evaluated on a new study design, which seems to be consistent with the general concern that the performance of newly proposed methods is over-optimistic \citep{buchka2021optimistic,norel2011self, boulesteix2013plea}.	Although a \lq\lq sample size'' of four and the re-evaluation of the methods on only one new design does not allow generalization of these results (neither for the considered methods nor for methodological research in general), the experiment provides insights into the mechanisms that can lead to performance differences between original and subsequent studies. In the following, we will discuss four of these mechanisms, which have either been addressed frequently in the literature or are rarely mentioned in literature but seem to have been present in our experiment. In addition, we point to possible solutions that can help to avoid large performance discrepancies between original and subsequent studies. \\

\noindent\textbf{Overfitting of study design to method}\space\space  
Our experiment illustrated the many degrees of freedom existing in the assessment of a method's performance. This flexibility can tempt researchers to choose the study design in favor of their proposed method. This may happen both at the planning stage when researchers primarily select a study design in which their method is expected to perform well (e.g., leaving competing methods at their default parameters), and after seeing the results when they add and/or omit certain design components (e.g., simulation parameters or evaluation criteria;
\citealp{ullmann2022over,Niessl2021,pawel2022pitfalls}). 
Focusing on advantageous designs at the planning stage is not necessarily a questionable research practice but becomes problematic if not clearly stated. Changing the study design \textit{after} seeing the results may be legitimate in some cases as far as it is transparently reported, for example if the originally chosen evaluation criterion turns out to behave inadequately for all methods. But changing the study design {\it is} bad practice if it is performed in a cherry-picking fashion, i.e. excluding or including results depending on whether they convey the expected message or not.  The ``overfitting'' of the study design to the method increases the risk of obtaining different, less optimistic conclusions in a subsequent comparison study in which the authors have less incentives to present the corresponding method in a favorable light.
 
As already noted by \cite{Simmons2011} in the context of applied research, such optimizations most often do not reflect malicious intent. Instead, they are usually the result of self-serving interpretations of ambiguity convincing honest researchers that the decisions (in our case, regarding the study design) matching their expectations and hopes are the most appropriate ones for various other reasons. These mechanisms are certainly encouraged by publication pressure and publication bias \citep{boulesteix2017towards}. Selective reporting after seeing the results can be largely avoided by pre-registering study designs and documenting all changes that have to be made subsequently \citep{morris2019using,pawel2022pitfalls}. However, it does not prevent authors from selecting advantageous designs from the start when planning their study. This pitfall could be avoided by adapting the designs from previous studies conducted by different authors, however, this might not always be suitable to demonstrate all features of the method. 

For the papers considered in our study, we do not assume that any components regarding the data sets, competing methods or evaluation criteria have been optimized to make the corresponding method of interest appear better than it actually is. On the other hand, we cannot completely rule out this possibility, although it is especially unlikely for NEMO, which was evaluated using a study design adopted from a
previously conducted comparison study \citep{Rappoport2018}, similar to pre-registration where the design is fixed in advance. \\

\noindent\textbf{Overfitting of method to study design}\space\space
Just as the study design can be ``overfitted'' to the method of interest, the method of interest can also be ``overfitted'', i.e., over-optimized to the study design. This was already noted by \cite{jelizarow2010over} and \cite{ullmann2022over} with a focus on overfitting to the considered data sets.
Since method development is, in itself, an optimization process that usually consists of several improvements after seeing the performance results, it is difficult to determine the point where further optimization amounts to overfitting the method to the design used for performance assessment.  
Note that this issue not only concerns the method characteristics that are not intended to be changed by the user, but also the parameters that can be set by the method user and whose optimal values for specific applications might also be overfitted to the considered study design \citep{ullmann2022over,pawel2022pitfalls}. 

To avoid overfitting of the method of interest to the study design, evaluating the method extensively is recommended. This includes using a large number of data sets and/or simulation settings and several evaluation criteria as well as checking the robustness of the method with respect to small changes in the study design since this makes it more difficult for the method to be artificially optimized \citep{norel2011self,boulesteix2015ten,ullmann2022over,Niessl2021}. In principle, this is comparable to the classical context of regression where overfitting is less likely to occur if the number of observations is large.

Moreover, it may be helpful to re-evaluate newly developed methods using a different design after the termination of the trial-and-error process, which might yield slightly worse but likely more realistic performance results (in the sense that the performance discrepancy between original and subsequent papers decreases). Although previous literature usually focuses on evaluating the method on new data \citep{jelizarow2010over,norel2011self,ullmann2022over}, considering different competing methods and evaluation criteria could also be reasonable. To reduce the risk of choosing the new design in favor of the proposed method, one could apply the design of a previous study conducted by different authors. As discussed above, the design of a previous study might not be suited to present all features of the proposed method but this might be less relevant if the design is considered as an additional ``external validation design''. An external validation design could be, for example, the design of a neutral comparison study, or, similar to our experiment, a previously proposed method (e.g., a method that was included as  competing method). This procedure is only feasible without much additional effort if the authors of the previous paper have made the code for reproducing the results openly available and does not protect against systematic manipulation (e.g., modifying the method after seeing the results and thus consciously biasing the external validation).

When reading a paper, it is typically not possible to identify whether the method of interest has been overfitted to the design used for method development and, unless explicitly stated, if there are any settings that have been separated from the development process. This also applies to the papers included in our experiment, which do not have a corresponding statement. However, \mbcdeg is mainly based 
on an algorithm that was developed by different authors for a different analysis task (i.e., clustering of genes that have already been identified as differentially expressed), which means that this part of the method could not be overfitted to the design of \cite{Osabe2021}.\\

\noindent\textbf{Different levels of expertise}\space\space
While the mechanisms discussed above are mostly attributed to the non-neutrality of the authors proposing their new method, there are also other potential mechanisms leading to deteriorating performances in subsequent papers. One of them originates from the fact that, as already noted by \cite{Duin1996}, the performance of a method is not just dependent on the design it is evaluated on but also on the skill of the person who applies the method. The difference in performance between original and  subsequent papers may also be due to the lower expertise level of the subsequent authors whose parameter choice when applying the method on the new data is likely to be less optimal than the parameters that the authors of the method would have selected. Of course, the degree to which the performance deteriorates due to the lack of expertise may be different for each method \citep{boulesteix2017towards} and also depends on how much the new design in which the method is applied differs from the design of the original paper.  

As described in Section~\ref{section:challenge_omics} and~\ref{section:challenge_de}, we also faced the challenge of choosing appropriate method parameters when applying the methods of our experiment to the new data sets and we cannot rule out that these decisions might have led to a worse performance than if the authors of the original papers had chosen the parameters themselves. In the first example on cancer subtyping, we note that although the data sets in both papers had the same data type and originated from the same source, the authors of NEMO and PINSPlus might have set different parameters (including method specific pre-processing steps) for their respective method since the data sets have a different distribution of samples and omic variables (due to the different pre-processing steps and number of data sets). For example, the authors of \pinsplus might have normalized the data when applying it to the data sets of \nemo. The same applies to the differential expression analysis example, where we decided to set SFMEB’s parameters for the crossed simulation data as in the simulation setting of the original paper that seemed to be the most similar to the new simulation. It is possible that the authors of \sfmeb who are experts for this method might have used a different parameter setting. For \mbcdeg, we also cannot rule out that our low level of expertise has contributed to the deteriorating performance of the method. Although we evaluated different values for one parameter of \mbcdeg as a sensitivity analysis, we only did that to a limited extent and the considered values may still be suboptimal (for example, the authors did not specify how the parameter should be set in the presence of uniquely expressed genes, which are not considered in their simulation settings). It also has to be noted that we are non-expert users for many of the competing methods used for each paper, and for instance the performance of Consensus Clustering and iCluster+ (competing methods of \pinsplus) is certainly dependent on the expertise level of the user since the optimal number of clusters has to be specified manually based on different types of plots and is thus very subjective. However, the difference in expertise (i.e., comparing our expertise vs. the expertise of the authors of the four papers) is probably less drastic with regard to the competing methods than for the methods of interest, and is thus not of equal relevance. 

One possibility to avoid the systematic deterioration of performance in subsequent studies 
due to a lower level of expertise is to involve the authors of the method in the respective study \citep{boulesteix2017towards, morris2019using,pawel2022pitfalls}. This can be realized if they implement their method themselves, as done for example in the study by \cite{zapf2021meta} that involved the authors of all considered methods as co-authors or in benchmark studies that are organized as challenges such as the DREAM challenges (\url{https://dreamchallenges.org/}). Alternatively, the authors of a method can be contacted to make sure that their method is implemented correctly as done in the comparison study by \cite{herrmann2021large}.  However, while the authors of a method could potentially be involved in the majority of comparison studies that assess their method, they will not be able to verify the correct implementation of their method in every \textit{applied} study. While there is value in studying the performance of a method when used by an expert, it might thus be even more important to assess the performance when it is applied by non-experts \citep{Duin1996, boulesteix2017towards}, as we did in this experiment. Note, however, that even among the non-experts of a method, there are different levels of expertise - or a different willingness to gain expertise by getting more familiar with the method (which may apply in particular to authors that use the method as competitor for their own method).

In general, it might thus be advisable for authors to make the performance of their method less dependent on user expertise, e.g., by focusing on concrete guidelines on how to choose optimal parameter values in different applications or by implementing automated parameter selection. The latter, although not always feasible, would protect against the above-mentioned tendency to leave method parameters of competing methods at default values. Moreover, reporting the robustness of the method performance with respect to different parameter values (as done by all four paper considered in the experiment) allows method users to gain understanding on which parameters need to be carefully specified \citep{ullmann2022over}. 
Of course, reducing the effect of different levels of expertise also requires efforts from the authors of subsequent papers who need to consider the available guidelines and information on how to set the method parameters.\\

\noindent\textbf{Different fields of application}\space\space
An insight we gained from the experiment that seems to be rarely addressed in the literature but plays an important role for the optimistic performance evaluation of newly proposed methods is related to the appropriate field of application of a method and its individual strengths within this field.  If a method performs worse in a subsequent paper, this can indeed be due to the mutual overfitting of method and design or the lack of expertise, as discussed above. However, the deteriorating performance could also be explained by the fact that the field of application of the subsequent study does not exactly match the field of application the method is intended for. Unfortunately, our experiment suggests that it often hard to assess if this is the case.

For example, although NEMO and PINSPlus obviously have the same \textit{general} field of application (i.e., cancer subtyping using multi-omic data), it was clear that \pinsplus, in contrast to NEMO, is not intended to be used on partial multi-omic data sets (i.e., data sets where some patients do not have any measurements for one or more omic data type), which is why we excluded them from our experiment. On the other hand, \pinsplus was initially (i.e., in its original paper) only evaluated based on its ability of finding subtypes that have significantly different survival while NEMO was additionally assessed based on the enrichment of certain clinical variables such as tumor stage. We did not exclude the clinical enrichment criterion, although it could be argued that \pinsplus is only intended for applications where it is relevant to find subtypes with different survival. 

Similarly, in the differential expression analysis example, we excluded the three-group simulated data used to assess the performance of \mbcdeg in the original paper since the authors of \sfmeb did not explicitly mention that their method is intended for this type of application.
On the other hand, we did not exclude the settings without biological replicates (i.e., $n_{obs}=$1 in each group) used by the authors of \sfmeb from our experiment although the authors of \mbcdeg did not explicitly state that settings without biological replicates belong to \mbcdeg's field of application (and other popular methods such as edgeR and DESeq2 are explicitly not intended for this setting). Moreover, it is not clear whether \mbcdeg can be applied in settings with uniquely expressed genes (i.e., genes with zero counts in one condition), which were included in most settings used to evaluate \sfmeb. 

These examples show that it is often not clear for method users what the method's exact field of application is, which consequently makes decisions on whether it is appropriate to apply the method to a new design more difficult and subjective.
On the other hand, authors proposing a new method cannot be expected to provide an exact definition of the method’s appropriate field of application that accounts for every imaginable design, and some authors explicitly state that the method simply requires more evaluation in certain designs to assess whether they belong to the method's appropriate field of application. For example, the authors of \mbcdeg mention that their method still needs to be evaluated on additional simulation frameworks and real data with different experimental settings and organisms. 

In general, authors proposing a new method should thus try to study and define its field of application as comprehensively as possible, while authors applying the method in a subsequent study should carefully check whether the inclusion of the method is appropriate. 

An issue related to the field of application is that methods often have specific strengths or features \textit{within} their field of application, which is typically reflected by the design and not problematic if reported transparently (as discussed above). 
However, the method’s strengths and special features may not be highlighted to the same extent through the design of the subsequent study (which may be for instance selected to highlight the strengths of a different method), thus leading to a deteriorating performance. 

We also observed this in our experiment. As mentioned above, a special feature of NEMO is that it can handle missing values in the omic data. However, this feature does not come into play in the study design of PINSPlus, which cannot handle missing values (so that its authors did not consider designs with missing data). Notably, NEMO outperformed the competing methods in the original paper even more clearly for the data sets with missing values than for the full data sets, and although NEMO showed good performance in the design of PINSPlus, its performance might have been even better if the crossed design had also included data sets with missing values. 

In the differential expression example, the authors of \sfmeb emphasize its strength of not requiring data normalization, which is an essential step for most other methods that can mislead downstream analysis if not done correctly. The authors of \sfmeb  include several data settings where normalization can be error-prone, such as heterogeneous data sets with clearly different fold-changes between the conditions. This special strength is however not relevant for the settings of \mbcdeg that are included in our experiment, which may have also led to SFMEB’s deteriorating performance.

In contrast to the mismatch regarding the appropriate field of application discussed above, it is no necessarily inappropriate if a subsequent study disregards the strengths of a method, but it should be ideally mentioned. 
Note that the discussed mechanisms can also be applied to the competing methods of the original and subsequent papers, whose field of application and specific strengths might be more or less reflected by the study design.

\section{Conclusion} \label{section:conclusion}
Based on the insights gained from the cross-design validation experiment, we conclude that while the discrepancy between original and subsequent studies assessing the performance of a method may be, in part, attributed to the non-neutrality of the method's authors, there are also other mechanisms related to different levels of expertise and fields of application that can contribute to a deteriorating method performance. 
It is important that both the authors proposing a method and the authors applying the method in a subsequent study acknowledge and counteract these mechanisms. Moreover, a minimum requirement for all papers proposing and/or comparing methods should be to openly provide the code and, if possible, data to reproduce the results. This does not guarantee but at least facilitates the detection of potential over-optimistic statements in the original papers and non-appropriate use of the methods in subsequent papers. In the long run, these efforts will increase the reliability of studies proposing new methods.

\vspace*{1pc}
\noindent {\bf{Acknowledgements}}
 This work was supported by the German Federal Ministry of Education and Research
(01IS18036A) and by the German Research Foundation (BO3139/4-3, BO3139/7-1,
BO3139/6-2) to ALB. The authors of this work take full responsibilities for its content.
The authors thank Milena Wünsch for helpful comments and Anna Jacob for language correction.

\vspace*{1pc}

\noindent {\bf{Conflict of Interest}} The authors have declared no conflict of interest

\vspace*{1pc}

\noindent {\bf{Data availability statement}} The \texttt{R} code and data to reproduce all results are openly available at \url{https://doi.org/10.6084/m9.figshare.20754028.v1}

\section*{Appendix}
\appendix 
\section{Additional information: Cancer subtyping using multi-omic data}

\subsection{Data pre-processing} \label{appendix:omics:prepro}
 \begin{table}[ht]  \small
\caption{Data pre-processing steps performed in the original papers of \pinsplus and \nemo. Competing methods not included in the experiment are indicated by asterisks (*).  In case of design-implementation-gaps, the information shown in the table refers to the code for reproducing the results.}
\label{tab:design_multiomics_supp}
\begin{tabularx}{\textwidth}{|X|L{5.5cm}|L{5.5cm}|}
 \hline  
& \textbf{PINSPlus} \citep{Nguyen2019} & \textbf{NEMO} \citep{Rappoport2019}\\ 
\hline 
\raggedright Missing values in omic data &
\begin{tabitemize}
\item 	No information on removal and imputation of missing values
\item 	Only use patients with observations in all three omics
\end{tabitemize} &
\begin{tabitemize}
\item 	Remove patients and omic variables with more than 20\% missing values; impute remaining missing values with k nearest neighbor imputation
\item 	Only use patients with observations in all three omics 
\end{tabitemize}\\
\hline
\raggedright Missing values in survival data &
Only include patients with non-missing survival information &
Set missing survival times and death info to 0 (these patients can still be used for clinical enrichment)\\ \hline
\raggedright Remove sample types not corresponding to “Primary solid Tumor” (e.g., ``Metastatic'')&
No information &
Yes, except for LAML and SKCM data set \\
\hline 
\raggedright Omic variable pre-processing for all methods &
\begin{tabitemize}
\item $\mathrm{log}_2$ transformation for gene expression and miRNA expression 
\end{tabitemize}&
\begin{tabitemize}
\item 	$\mathrm{log}$ transformation for gene expression and miRNA expression 
\item	In all three omics, remove variables with zero variance
\item   For methylation data, select 5000 variables with maximal variance
\end{tabitemize} \\
\hline
\raggedright Method-specific pre-processing &
For all three types of omic data: \vspace{0.3cm}
\begin{tabitemize}
\item 	Normalize variables (mean 0, standard deviation 1): SNF, Consensus Clustering
\item	Subtract median after normalization: Consensus Clustering
\item	Select 2000 variables with max. median absolute deviation: \iclusterplus
\item	Remove variables with zero variance: \iclusterplus 
\end{tabitemize} &
For all three types of omic data:\vspace{0.3cm}
\begin{tabitemize}
\item 	Normalize variables (mean 0, standard deviation 1) : $k$-means, spectral clustering, SNF, MCCA, NEMO, *rMKL-LPP
\item	Select 2000 variables with highest variance: $k$-means, Spectral, MCCA, *MultiNMF
\end{tabitemize}\\
\hline

\end{tabularx}
\end{table}

\begin{table}[ht]
\caption{Number of patients and omic variables (gene expression, methylation,miRNA expression) after all pre-processing steps (except method-specific pre-processing) have been performed.}
\label{tab:omicsinput}
\centering \small
\begin{tabular}{|l|r|r|r|r|r|r|r|r|}
\hline 
& \multicolumn{4}{l|}{\textbf{PINSPlus} \citep{Nguyen2019}} &\multicolumn{4}{l|}{ \textbf{NEMO} \citep{Rappoport2019}} \\
  \hline
 \thead[l]{Data set} & \thead[l]{Patients} & \thead[l]{Gene \\expression} & \thead[l]{Methylation}& \thead[l]{MiRNA\\expression}  & \thead[l]{Patients} & \thead[l]{Gene \\expression}  & \thead[l]{Methylation\\}& \thead[l]{MiRNA\\expression} \\ 
  \hline
 BRCA & 622 & 239322 & 363763 & 2588 & 621 & 20226 & 5000 & 891 \\ 
 COAD & 220 & 239322 & 374946 & 30771 & 220 & 19991 & 5000 & 613 \\ 
 GBM & 273 & 12042 & 22833 & 534 & 274 & 12042 & 5000 & 534 \\ 
 KIRC & 124 & 17974 & 23165 & 590 & 183 & 20087 & 5000 & 796 \\ 
LAML & 164 & 16818 & 22288 & 552 & 170 & 19938 & 5000 & 558 \\ 
  LIHC & 366 & 73599 & 369193 & 540 & 367 & 20153 & 5000 & 852 \\ 
 LUSC & 110 & 12042 & 23348 & 706 & 341 & 20237 & 5000 & 878 \\ 
 OV & 286 & 239322 & 21675 & 705 & 287 & 20174 & 5000 & 616 \\ 
 SARC & 257 & 20531 & 374752 & 1046 & 257 & 20221 & 5000 & 838 \\ 
 SKCM & 439 & 20531 & 373814 & 586 & 448 & 20226 & 5000 & 901 \\ 
   \hline
\end{tabular}
\end{table} 
As displayed in Table~\ref{tab:design_multiomics_supp}, \ngu and \rapp use different data pre-processing steps. This includes for example the way missing data are handled or omic variables are selected and normalized.
For SNF, the only method that is considered as competing method in both papers, each omic variable (for all three omic types) is normalized to have a mean of 0 and a standard deviation of 1 in \ngu, while \rapp also remove omic variables with zero variance and select the 5000 variables with the highest variance for the methylation data (this is done for all methods in \rapp).
Note that the information on pre-processing shown in Table \ref{tab:design_multiomics_supp} is based on the published code and, as far as early pre-processing that generates the data provided by the authors is concerned, on the text in the papers. 
This means that there could have been more pre-processing steps that are not reported. Table \ref{tab:omicsinput} shows the resulting number of patients and omic variables for \ngu and \rapp after applying the pre-processing steps.

As stated in Section \ref{section:experiment_term}, we consider all pre-processing steps that are performed for \textit{all} methods as belonging to the data component and method-specific pre-processing steps as belonging to the respective methods. However, some refinements are necessary when crossing the designs. More specifically, we note that iClusterBayes and LRACluster (competing methods of \rapp) have very long runtimes when run on the data sets of \ngu. This is because \ngu do not perform any variable selection as a general pre-processing step (only for \iclusterplus). Hence, when running iClusterBayes and LRACluster on the data from \ngu, we select the 2000 omic variables with the highest variance for each omic data type as it is done for $k$-means, spectral clustering, MCCA, and MultiNMF in \rapp.

\subsection{Reproducibility issues for two competing methods}\label{appendix:omics:fail}
We have to exclude two competing methods of \nemo (rMKL-LPP and MultiNMF) from the experiment. In the README file accompanying the code of \cite{Rappoport2018}, the authors state that reproducing the results of rMKL-LPP requires the source code of the method, which they report is only available on request from the authors of rMKL-LPP. It seems that the method can also be run on a web server by now (\url{www.web-rMKL.org}), which, however, is not available at the time of writing  (last checked in August 2022). Moreover, we have to exclude MultiNMF since running the \texttt{R} code provided by \cite{Rappoport2018} (and thus by \rapp) requires that the user inserts MATLAB commands, which we are not able to specify correctly. Note that \cite{Tepeli2020} were also not able to reproduce the results of MultiNMF shown in \cite{Rappoport2018}

\subsection{Approximation-based vs. permutation-based $p$-values} \label{appendix:omics:permutation}
\begin{figure}[ht]
     \centering
     \includegraphics[width=0.6\textwidth]{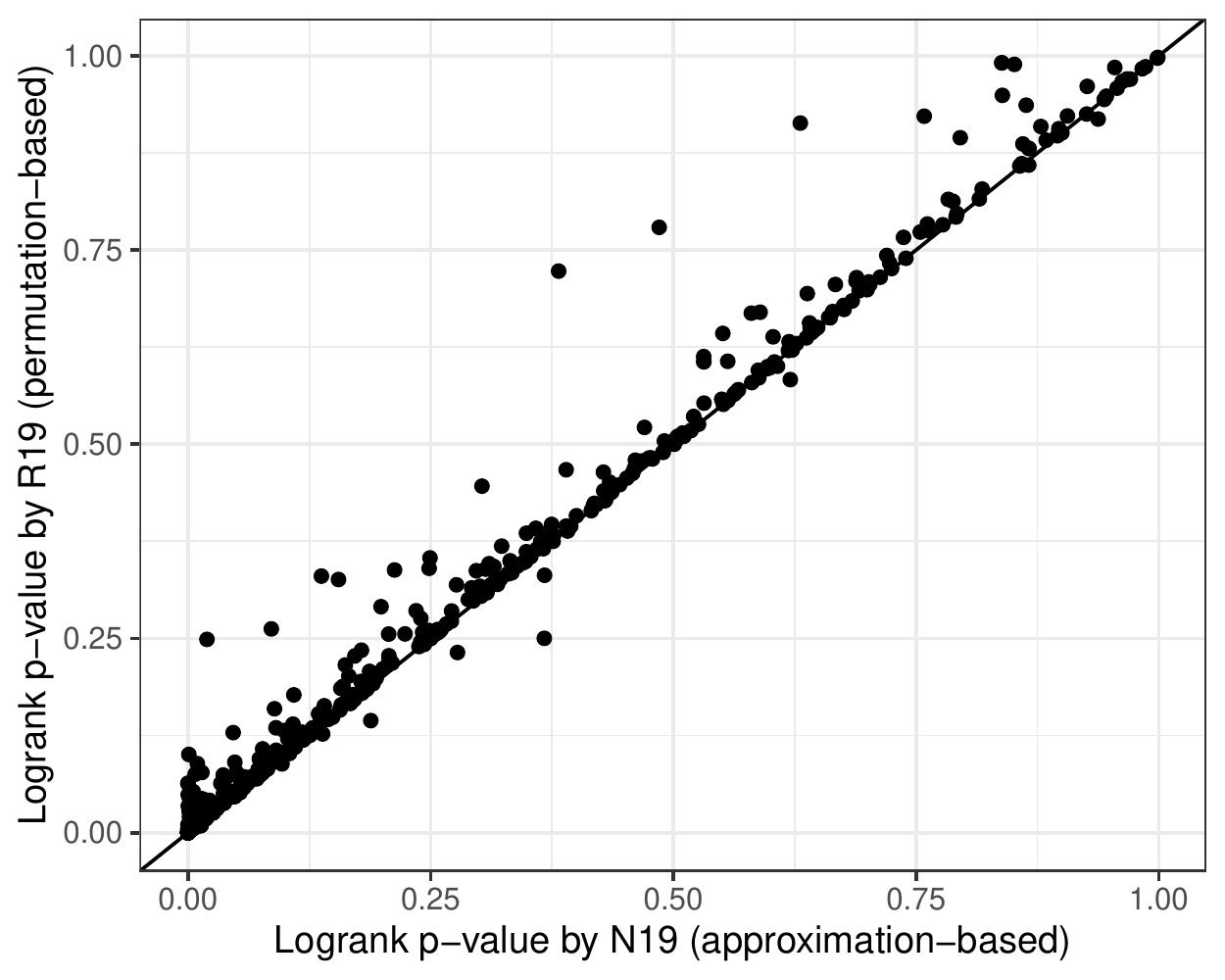}
     \caption{Comparison of approximation-based and permutation-based $p$-values. Each point refers to the logrank $p$-value of a method when applied to a data set. All methods and data sets considered by \ngu and \rapp are included, resulting in 528 points ($12$ methods $\times$ 44 data sets).}
     \label{fig:appendix:pvalue}
 \end{figure}

 \cite{Rappoport2018} note that the $\rchi^2$ distribution assumed for the test statistics of the logrank, the  $\rchi^2$,  and  the Kruskal-Wallis test is not an accurate approximation for small sample sizes and unbalanced cluster sizes, especially for large values of the test statistic. Hence, \cite{Rappoport2018} (and thus also \rapp) estimate the $p$-values using permutation procedures, i.e., they randomly permute the cluster labels and calculate empirical $p$-values as the fraction of permutations for which the test statistic is greater or equal than the test statistic yielded by the original clustering. \cite{Rappoport2018} report that they observed large differences between approximation-based (i.e., assuming $\rchi^2$ distribution) and permutation-based  $p$-values, with the former yielding increased type 1 errors. They conclude that at least for TCGA data sets, analyses that use approximation-based $p$-values might not be valid. In our experiment, the approximation-based $p$-values are indeed generally smaller, as can be seen from Figure \ref{fig:appendix:pvalue}
 
\subsection{Reproduced performance results of \pinsplus and \nemo for each data set} \label{appendix:omics:origresults}
\begin{table}[ht]
\centering \small
\caption{Reproduced performance results (logrank $p$-values) of \pinsplus and its competing methods for each data set based on the original study design by \ngu.} 
\label{appendix:tab:pinsplus}
\begin{tabular}{|r|l|l|l|l|l|}
  \hline
 & Data set & \pinsplus & CC & SNF & iCluster+ \\ 
  \hline
1 & KIRC & 6e-05 & 0.118 & 0.691 & 0.058 \\
  2 & GBM & 8.7e-05 & 0.014 & 0.021 & 0.103 \\ 
  3 & LAML & 0.00087 & 0.292 & 0.002 & 0.083 \\ 
  4 & LUSC & 0.008 & 0.688 & 0.087 & 0.224 \\ 
  5 & BLCA & 0.019 & 0.089 & 0.109 & 0.17 \\ 
  6 & HNSC & 0.046 & 0.428 & 0.366 & 0.364 \\ 
  7 & LIHC & 0.03 & 0.622 & 0.334 & 0.072 \\ 
  8 & STAD & 0.002 & 0.428 & 0.041 & 0.434 \\ 
  9 & THYM & 0.013 & 0.139 & 0.097 & 0.24 \\ 
  10 & GBMLGG & 7.5e-17 & 0.00052 & 4.8e-14 & 5.4e-14 \\ 
  11 & LGG & 7.7e-25 & 2e-06 & 1.6e-14 & 2.7e-14 \\ 
  12 & PAAD & 0.00025 & 0.013 & 0.00074 & 0.00063 \\ 
  13 & SKCM & 0.048 & 0.604 & 0.478 & 0.108 \\ 
  14 & COADREAD & 0.003 & 0.946 & 0.66 & 0.178 \\ 
  15 & UCEC & 0.001 & 0.105 & 0.018 & 0.619 \\ 
  16 & CESC & 0.03 & 0.376 & 0.51 & 0.201 \\ 
  17 & COAD & 0.001 & 0.419 & 0.128 & 0.884 \\ 
  18 & BRCA & 0.007 & 0.008 & 0.119 & 0.046 \\ 
  19 & STES & 0.007 & 0.301 & 0.157 & 0.46 \\ 
  20 & KIRP & 1.1e-09 & 0.367 & 0.005 & 0.013 \\ 
  21 & KICH & 0.028 & 0.955 & 0.701 & 0.788 \\ 
  22 & UVM & 0.00075 & 0.005 & 0.00017 & 0.003 \\ 
  23 & ACC & 0.007 & 0.014 & 4.3e-05 & 0.00071 \\ 
  24 & SARC & 0.03 & 0.148 & 0.044 & 4e-04 \\ 
  25 & MESO & 0.00073 & 0.272 & 0.00042 & 0.00022 \\ 
  26 & READ & 0.649 & 0.737 & 0.762 & 0.249 \\ 
  27 & UCS & 0.458 & 0.207 & 0.859 & 0.983 \\ 
  28 & OV & 0.319 & 0.859 & 0.445 & 0.062 \\ 
  29 & ESCA & 0.33 & 0.791 & 0.392 & 0.16 \\ 
  30 & PCPG & 0.866 & 0.938 & 0.332 & 0.55 \\ 
  31 & LUAD & 0.099 & 0.926 & 0.501 & 0.118 \\ 
  32 & PRAD & 0.349 & 0.638 & 0.475 & 0.879 \\ 
  33 & THCA & 0.166 & 0.64 & 0.62 & 0.111 \\ 
  34 & TGCT & 0.531 & 0.758 & 0.838 & 0.58 \\ 
   \hline
\end{tabular}

\end{table}

\begin{table}[ht]
\centering \small
\caption{Reproduced performance results (number of enriched clinical variables / $-\mathrm{log}_{10}$ logrank $p$-values) of \nemo and its competing methods for each data set based on the original study design by \rapp.} 
\label{appendix:tab:nemo}
\begin{tabular}{|r|l|l|l|l|l|l|l|l|l|}
  \hline
 & Data set & K-Means & Spectral & LRACluster & PINS & SNF & MCCA & iClusterBayes & NEMO \\ 
  \hline
1 & LAML & 1/2.9 & 1/1.9 & 1/2 & 1/1.1 & 1/2.9 & 1/1.4 & 1/0.9 & 1/2.1 \\ 
  2 & BRCA & 0/0.6 & 2/1.6 & 4/1.3 & 1/1.2 & 2/1 & 0/3.2 & 3/0.2 & 3/1.4 \\ 
  3 & COAD & 0/0 & 0/0.2 & 0/0.5 & 0/0 & 0/0.2 & 1/0.3 & 0/0.2 & 0/0.2 \\ 
  4 & GBM & 2/2.3 & 2/2.3 & 1/1.4 & 1/3.6 & 1/4.2 & 1/1.9 & 0/1 & 1/1.9 \\ 
  5 & KIRC & 0/0.2 & 0/0.3 & 0/4.5 & 0/1.8 & 1/2.1 & 1/3.8 & 1/2 & 1/1.2 \\ 
  6 & LIHC & 1/0.2 & 2/0.4 & 0/0.8 & 2/2 & 2/0.2 & 2/0.9 & 2/1 & 3/3.3 \\ 
  7 & LUSC & 1/0.2 & 2/0.3 & 1/0.9 & 0/0.3 & 0/0.6 & 0/0.4 & 2/0.6 & 0/0.4 \\ 
  8 & SKCM & 2/0.6 & 2/0.9 & 3/2.7 & 1/2.8 & 1/0.6 & 2/4.3 & 3/4.4 & 3/3.9 \\ 
  9 & OV & 1/0.1 & 1/0.8 & 1/0.6 & 0/0 & 0/0.2 & 1/0.7 & 0/0 & 1/0.1 \\ 
  10 & SARC & 2/1.3 & 2/1.3 & 2/1 & 2/1.2 & 2/2.1 & 2/0.6 & 2/0.8 & 2/1.8 \\ 
   \hline
\end{tabular}
\end{table}

  Table~\ref{appendix:tab:pinsplus} and \ref{appendix:tab:nemo} display the reproduced results of \nemo and \pinsplus for each data set. 
 
\subsection{Comparison of SNF implementations}
\begin{figure}[ht]
     \centering
     \includegraphics[width=1\textwidth]{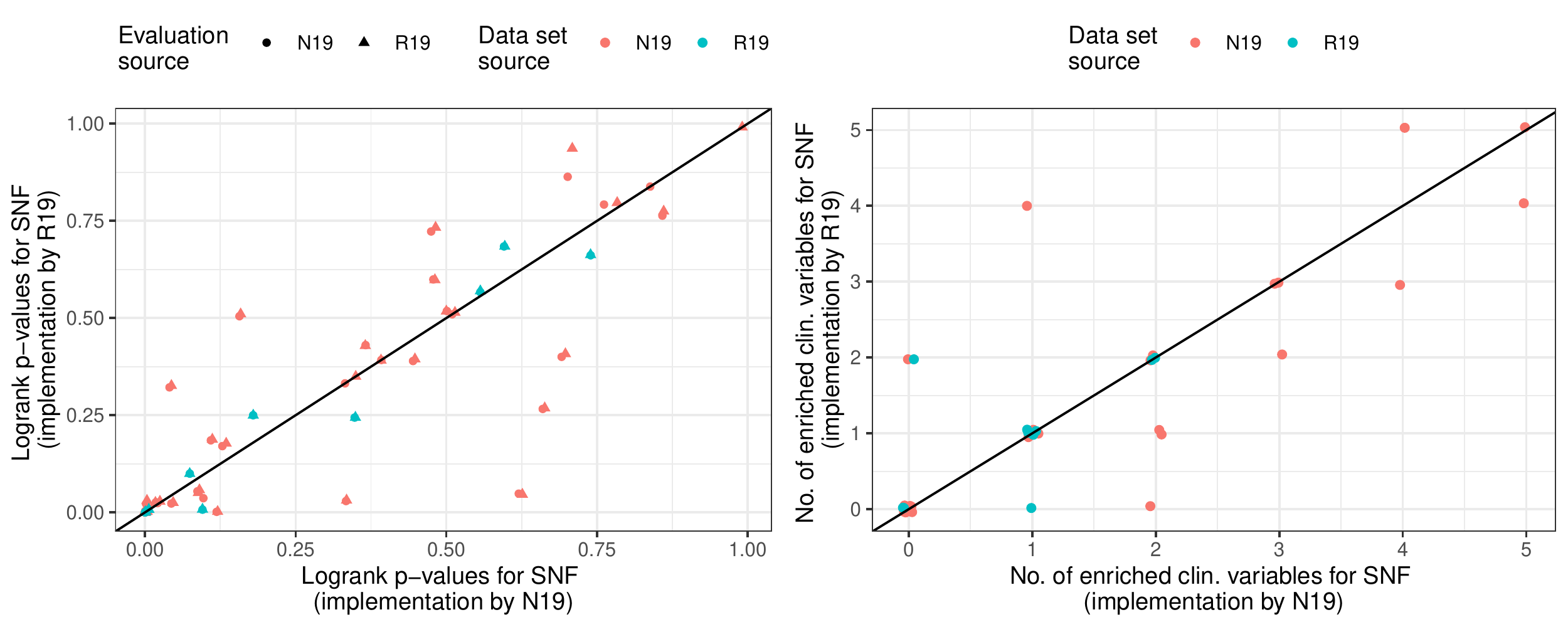}
     \caption{Logrank $p$-values and number of enriched clinical variables resulting from the two different SNF implementations specified by \ngu and \rapp. The left panel includes 88 points, representing the two different $p$-value estimation procedures on all 44 (34 +10) data sets considered by \ngu and \rapp. The right panel includes 44 points since the $p$-values for clinical enrichment are only calculated based on permutation tests.}
     \label{fig:appendix:snf}
 \end{figure}

The cancer subtyping method SNF is used as competing method for both \pinsplus and \nemo. However, \ngu and \rapp set different method parameters for SNF. 
Figure \ref{fig:appendix:snf} shows the logrank $p$-values and number of enriched clinical variables resulting from the two different implementations, revealing a considerable but non-systematic performance difference. 
 
 \clearpage
\section{Additional information: Differential gene expression analysis}

\subsection{Incorrect AUC calculation}\label{appendix:de:auc}
\zhou use the \texttt{pROC} \citep{procpackage} to calculate the AUC. 
\zhou and \osabe use different \texttt{R} packages for calculating the AUC, namely \texttt{ROC} \citep{rocpackage} and \texttt{pROC} \citep{procpackage}, respectively. In the \texttt{pROC} package, the function that calculates the ROC curve (\texttt{roc}) takes the argument \texttt{direction}, which determines whether values higher or lower than the threshold should be considered as cases (i.e., DE genes in this context). Per default, the package sets the direction \textit{automatically} according to the medians of the predicted values (see argument \texttt{direction} in the \texttt{roc} function of the \texttt{pROC} manual), which implies that the ROC curves are biased towards higher AUC values if the \texttt{direction} argument is not set explicitly. More precisely, this means that if the automatically defined \texttt{direction} argument is not correct, the resulting AUC will be 1 minus the correct AUC. It seems as if \zhou were not aware of this unfortunate default option since they did not explicitly specify the \texttt{direction} argument, potentially leading to incorrect AUC values.

\subsection{Different edgeR implementations}\label{appendix:de:edger}
While the edgeR implementation used by \osabe corresponds to one of the edgeR standard workflows, \zhou use three different versions of edgeR, of which only one can be considered as standard edgeR workflow (still using a slightly different version than \osabe). In six simulation settings, \zhou only use an edgeR-like implementation, which is not based on the negative binomial distribution that is usually considered for edgeR but on the Poisson distribution (presumably, this is done because the counts in these settings are generated using Poisson distribution). Since \zhou also include settings with no biological replicates (i.e., $n=1$ in each group) where edgeR results in an error, they instead use a testing procedure involving a binomial test.  While there are in fact several options suggested by the edgeR user manual (Section 2.12 - \textit{What to do if you have no replicates}) for settings with no biological replicates (although it is stated that these options are not ideal), these do not include the procedure used by \zhou. Instead, it is mentioned as an option for technical replicates (i.e., repeated measurements of the same sample that represent independent measures of the random noise associated with protocols or equipment; \citealp{Blainey2014}).

\subsection{Sensitivity analysis of \mbcdeg}\label{appendix:de:sa}
\begin{figure}[ht]
     \centering
     \includegraphics[width=1\textwidth]{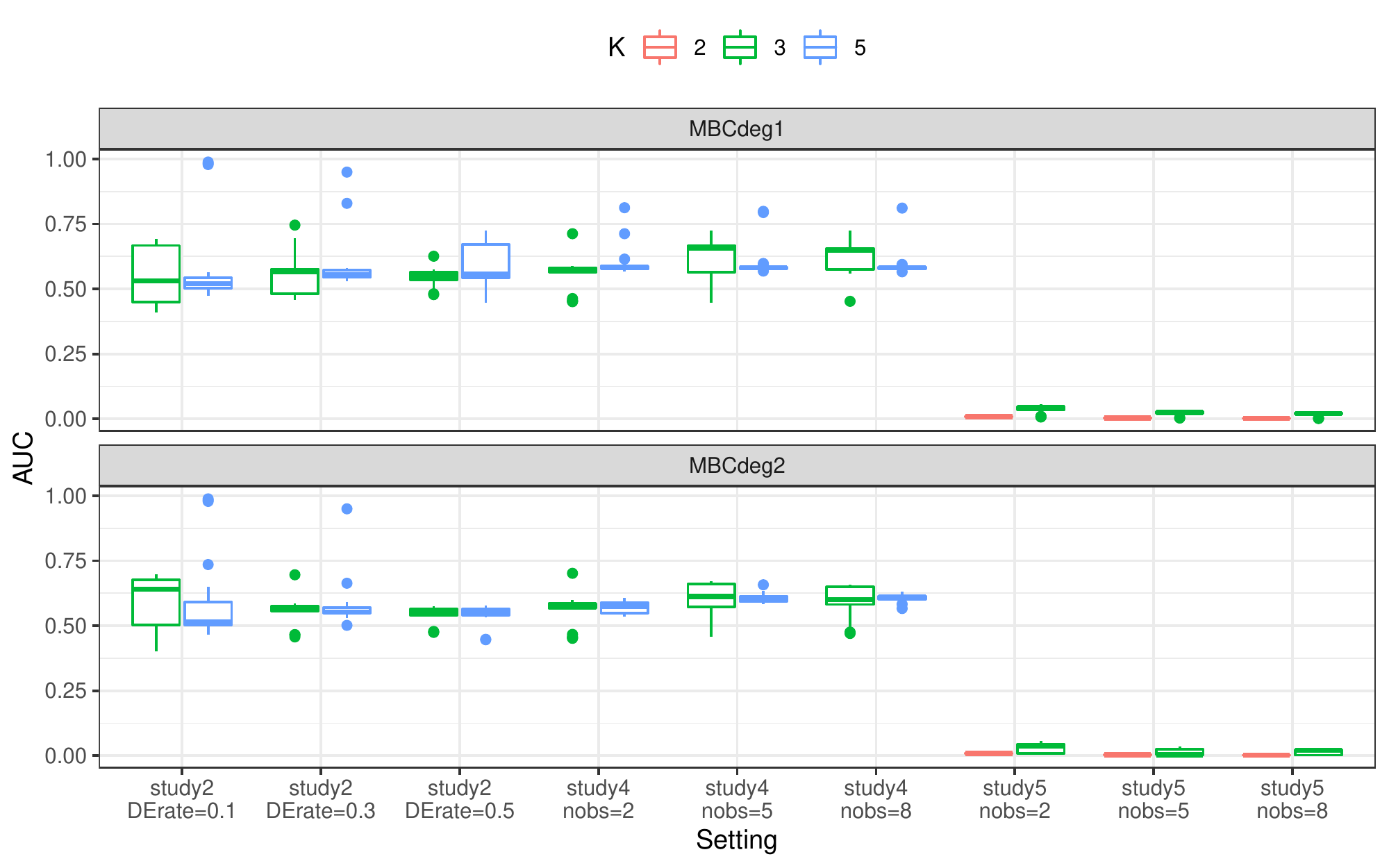}
     \caption{Performance results for \mbcdega and \mbcdegb when using different values for $K$.}
     \label{fig:appendix:sa}
 \end{figure}

The main parameter of \mbcdeg (which is based on a clustering algorithm) is the number of clusters $K$ to be found by the method.	The number of clusters does not have a default value and is set to $K=3$ by \osabe in the simulation settings that we reproduce in our experiment. This reflects the assumption that there are three gene expression patterns: non-DE genes, DE genes up-regulated in group 1, and DE genes up-regulated in group 2 (where up-regulated in group $j$ again means having higher expression in group $j$).
However, \osabe note that for settings where genes that are up-regulated in one group show different degrees of differential expression (i.e., fold-changes), allowing \mbcdeg to generate a higher number of clusters could lead to more accurate results. This could apply to the settings of study 2 and 4 considered in \zhou, which consist of two data sets with two different $\mathrm{log}_2$ fold-changes (i.e., 2 and 3). As a sensitivity analysis, we thus set $K = 5$ for these settings, reflecting non-DE genes and the two different degrees of differential expression for both groups, which however does not result in higher AUC values (see Figure~\ref{fig:appendix:sa}). Moreover, \osabe state that for settings where all DE genes are up-regulated in one group, the true number of clusters is actually $K=2$, reflecting non-DE genes and DE genes (all up-regulated in one group). Since this situation is present for the three settings of study 5 in \zhou, we also run \mbcdeg with $K=2$, which, however, does not lead to improved results (see Figure~\ref{fig:appendix:sa}). 

\subsection{Experiment results of \mbcdega}\label{appendix:de:mbcdeg1}
Figure \ref{fig:appendix:mbcdeg1} presents the performance ranks of \mbcdega, which, in contrast to \mbcdegb uses the default normalization algorithm. 
  \begin{figure}[ht]
     \centering
     \includegraphics[width=0.7\textwidth]{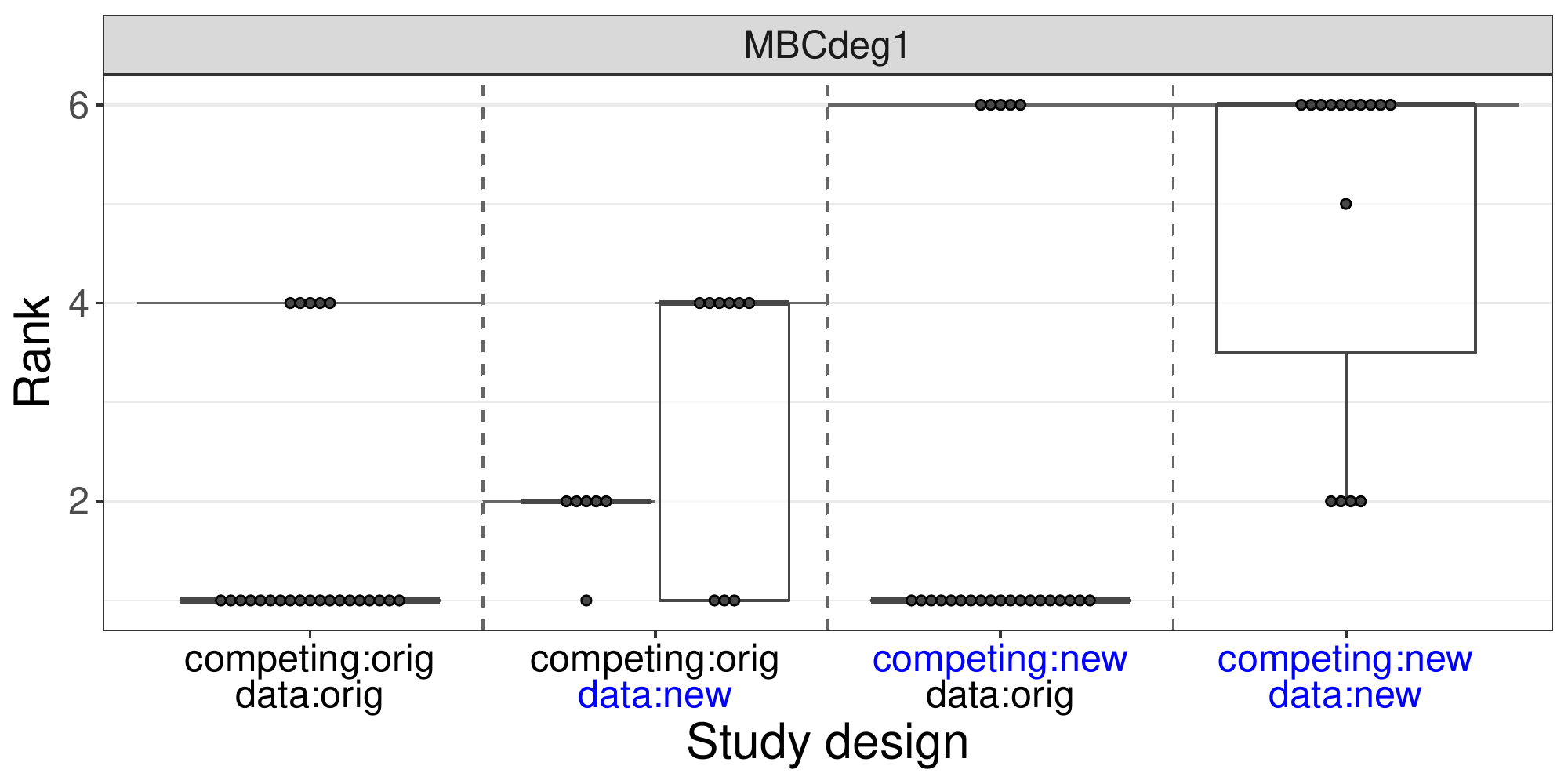}
     \caption{Performance ranks of \mbcdega based on data and competing methods that either correspond to the original \citep{Osabe2021} or crossed design \citep{Zhou2021}. Each boxplot consists of 15 or 24 ranks (additionally represented as black points), which corresponds to the number of data settings considered by  \cite{Zhou2021} and \cite{Osabe2021}, respectively. The ranks of \mbcdega in each data  setting are calculated based on the median AUC value across all simulation repetitions. The black lines correspond to the number of compared methods, i.e., the highest possible rank. Note that for one combination of data and competing methods, not all competing methods can be applied to all settings, which is why the number of compared methods varies between 2 and 4.}
     \label{fig:appendix:mbcdeg1}
 \end{figure}
 
 \clearpage
\clearpage

 \bibliography{bibliography}

\begin{thebibliography}{}

\bibitem[Anders and Huber, 2010]{Anders2010}
Anders, S. and Huber, W. (2010).
\newblock {Differential expression analysis for sequence count data}.
\newblock {\em Genome Biology}, 11:106.

\bibitem[Blainey et~al., 2014]{Blainey2014}
Blainey, P., Krzywinski, M., and Altman, N. (2014).
\newblock {Points of significance: Replication}.
\newblock {\em Nature Methods}, 11:879--880.

\bibitem[Boulesteix, 2015]{boulesteix2015ten}
Boulesteix, A.-L. (2015).
\newblock {Ten simple rules for reducing overoptimistic reporting in
  methodological computational research}.
\newblock {\em PLoS Computational Biology}, 11:e1004191.

\bibitem[Boulesteix et~al., 2013]{boulesteix2013plea}
Boulesteix, A.-L., Lauer, S., and Eugster, M.~J. (2013).
\newblock A plea for neutral comparison studies in computational sciences.
\newblock {\em PloS one}, 8:e61562.

\bibitem[Boulesteix and Strobl, 2009]{Boulesteix2009}
Boulesteix, A.-L. and Strobl, C. (2009).
\newblock Optimal classifier selection and negative bias in error rate
  estimation: an empirical study on high-dimensional prediction.
\newblock {\em BMC medical research methodology}, 9:85.

\bibitem[Boulesteix et~al., 2017]{boulesteix2017towards}
Boulesteix, A.-L., Wilson, R., and Hapfelmeier, A. (2017).
\newblock {Towards evidence-based computational statistics: Lessons from
  clinical research on the role and design of real-data benchmark studies}.
\newblock {\em BMC Medical Research Methodology}, 17:138.

\bibitem[Buchka et~al., 2021]{buchka2021optimistic}
Buchka, S., Hapfelmeier, A., Gardner, P.~P., Wilson, R., and Boulesteix, A.-L.
  (2021).
\newblock On the optimistic performance evaluation of newly introduced
  bioinformatic methods.
\newblock {\em Genome Biology}, 22:152.

\bibitem[Carey and Redestig, 2021]{rocpackage}
Carey, V. and Redestig, H. (2021).
\newblock {\em {ROC: utilities for ROC, with microarray focus}}.
\newblock R package version 1.62.0.

\bibitem[Duan et~al., 2021]{Duan2021}
Duan, R., Gao, L., Gao, Y., Hu, Y., Xu, H., Huang, M., Song, K., Wang, H.,
  Dong, Y., Jiang, C., Zhang, C., and Jia, S. (2021).
\newblock {Evaluation and comparison of multi-omics data integration methods
  for cancer subtyping}.
\newblock {\em PLoS Computational Biology}, 17:e1009224.

\bibitem[Duin, 1996]{Duin1996}
Duin, R. P.~W. (1996).
\newblock {A note on comparing classifiers}.
\newblock {\em Pattern Recognition Letters}, 17:529--536.

\bibitem[Herrmann et~al., 2021]{herrmann2021large}
Herrmann, M., Probst, P., Hornung, R., Jurinovic, V., and Boulesteix, A.-L.
  (2021).
\newblock Large-scale benchmark study of survival prediction methods using
  multi-omics data.
\newblock {\em Briefings in bioinformatics}, 22:bbaa167.

\bibitem[Hofner et~al., 2016]{Hofner2016}
Hofner, B., Schmid, M., and Edler, L. (2016).
\newblock {Reproducible research in statistics: A review and guidelines for the
  Biometrical Journal}.
\newblock {\em Biometrical Journal}, 58:416--427.

\bibitem[Jelizarow et~al., 2010]{jelizarow2010over}
Jelizarow, M., Guillemot, V., Tenenhaus, A., Strimmer, K., and Boulesteix,
  A.-L. (2010).
\newblock Over-optimism in bioinformatics: an illustration.
\newblock {\em Bioinformatics}, 26:1990--1998.

\bibitem[Klau et~al., 2020]{Klau2020}
Klau, S., Martin-Magniette, M.-L., Boulesteix, A.-L., and Hoffmann, S. (2020).
\newblock {Sampling uncertainty versus method uncertainty: A general framework
  with applications to omics biomarker selection}.
\newblock {\em Biometrical Journal}, 62:670--687.

\bibitem[Lohmann et~al., 2021]{lohmann2021s}
Lohmann, A., Astivia, O. L.~O., Morris, T., and Groenwold, R.~H. (2021).
\newblock It's time! 10+ 1 reasons we should start replicating simulation
  studies.

\bibitem[Love et~al., 2014]{Love2014}
Love, M.~I., Huber, W., and Anders, S. (2014).
\newblock {Moderated estimation of fold change and dispersion for RNA-seq data
  with DESeq2}.
\newblock {\em Genome Biology}, 15:550.

\bibitem[Monti et~al., 2003]{monti2003consensus}
Monti, S., Tamayo, P., Mesirov, J., and Golub, T. (2003).
\newblock Consensus clustering: a resampling-based method for class discovery
  and visualization of gene expression microarray data.
\newblock {\em Machine Learning}, 52:91--118.

\bibitem[Morris et~al., 2019]{morris2019using}
Morris, T.~P., White, I.~R., and Crowther, M.~J. (2019).
\newblock Using simulation studies to evaluate statistical methods.
\newblock {\em Statistics in medicine}, 38:2074--2102.

\bibitem[Nguyen et~al., 2019]{Nguyen2019}
Nguyen, H., Shrestha, S., Draghici, S., and Nguyen, T. (2019).
\newblock {PINSPlus: A tool for tumor subtype discovery in integrated genomic
  data}.
\newblock {\em Bioinformatics}, 35:2843--2846.

\bibitem[Nguyen et~al., 2017]{Nguyen2017}
Nguyen, T., Tagett, R., Diaz, D., and Draghici, S. (2017).
\newblock {A novel approach for data integration and disease subtyping}.
\newblock {\em Genome Research}, 27:2025--2039.

\bibitem[Nießl et~al., 2022]{Niessl2021}
Nießl, C., Herrmann, M., Wiedemann, C., Casalicchio, G., and Boulesteix, A.-L.
  (2022).
\newblock Over-optimism in benchmark studies and the multiplicity of design and
  analysis options when interpreting their results.
\newblock {\em WIREs Data Mining and Knowledge Discovery}, 12:e1441.

\bibitem[Norel et~al., 2011]{norel2011self}
Norel, R., Rice, J.~J., and Stolovitzky, G. (2011).
\newblock The self-assessment trap: can we all be better than average?
\newblock {\em Molecular Systems Biology}, 7:537.

\bibitem[Osabe et~al., 2021]{Osabe2021}
Osabe, T., Shimizu, K., and Kadota, K. (2021).
\newblock {Differential expression analysis using a model-based gene clustering
  algorithm for RNA-seq data}.
\newblock {\em BMC Bioinformatics}, 22:511.

\bibitem[Pawel et~al., 2022]{pawel2022pitfalls}
Pawel, S., Kook, L., and Reeve, K. (2022).
\newblock Pitfalls and potentials in simulation studies.
\newblock {\em arXiv preprint arXiv:2203.13076}.

\bibitem[Rappoport and Shamir, 2018]{Rappoport2018}
Rappoport, N. and Shamir, R. (2018).
\newblock {Multi-omic and multi-view clustering algorithms: review and cancer
  benchmark}.
\newblock {\em Nucleic Acids Research}, 46:10546--10562.

\bibitem[Rappoport and Shamir, 2019]{Rappoport2019}
Rappoport, N. and Shamir, R. (2019).
\newblock {NEMO: Cancer subtyping by integration of partial multi-omic data}.
\newblock {\em Bioinformatics}, 35:3348--3356.

\bibitem[Rigaill et~al., 2018]{Rigaill2018}
Rigaill, G., Balzergue, S., Brunaud, V., Blondet, E., Rau, A., Rogier, O.,
  Caius, J., Maugis-Rabusseau, C., Soubigou-Taconnat, L., Aubourg, S., Lurin,
  C., Martin-Magniette, M.-L., and Delannoy, E. (2018).
\newblock {Synthetic data sets for the identification of key ingredients for
  RNA-seq differential analysis}.
\newblock {\em Briefings in Bioinformatics}, 19:65--76.

\bibitem[Robin et~al., 2011]{procpackage}
Robin, X., Turck, N., Hainard, A., Tiberti, N., Lisacek, F., Sanchez, J.-C.,
  and Müller, M. (2011).
\newblock {pROC: an open-source package for R and S+ to analyze and compare ROC
  curves}.
\newblock {\em BMC Bioinformatics}, 12:77.

\bibitem[Robinson et~al., 2009]{Robinson2009}
Robinson, M.~D., McCarthy, D.~J., and Smyth, G.~K. (2009).
\newblock {edgeR: a Bioconductor package for differential expression analysis
  of digital gene expression data}.
\newblock {\em Bioinformatics}, 26.

\bibitem[Robinson and Oshlack, 2010]{Robinson2010}
Robinson, M.~D. and Oshlack, A. (2010).
\newblock {A scaling normalization method for differential expression analysis
  of RNA-seq data}.
\newblock {\em Genome Biology}, 11:R25.

\bibitem[Seyednasrollah et~al., 2013]{Seyednasrollah2013}
Seyednasrollah, F., Laiho, A., and Elo, L.~L. (2013).
\newblock {Comparison of software packages for detecting differential
  expression in RNA-seq studies}.
\newblock {\em Briefings in Bioinformatics}, 16:59--70.

\bibitem[Silberzahn et~al., 2018]{silberzahn2018many}
Silberzahn, R., Uhlmann, E.~L., Martin, D.~P., Anselmi, P., Aust, F., Awtrey,
  E., Bahn{\'\i}k, {\v{S}}., Bai, F., Bannard, C., Bonnier, E., et~al. (2018).
\newblock Many analysts, one data set: Making transparent how variations in
  analytic choices affect results.
\newblock {\em Advances in Methods and Practices in Psychological Science},
  1:337--356.

\bibitem[Simmons et~al., 2011]{Simmons2011}
Simmons, J.~P., Nelson, L.~D., and Simonsohn, U. (2011).
\newblock {False-positive psychology: Undisclosed flexibility in data
  collection and analysis allows presenting anything as significant}.
\newblock {\em Psychological Science}, 22:1359--1366.

\bibitem[Sonabend et~al., 2022]{sonabend2022avoiding}
Sonabend, R., Bender, A., and Vollmer, S. (2022).
\newblock Avoiding c-hacking when evaluating survival distribution predictions
  with discrimination measures.
\newblock {\em Bioinformatics}, page btac451.

\bibitem[Soneson, 2014]{compcodeR}
Soneson, C. (2014).
\newblock {compcodeR - an R package for benchmarking differential expression
  methods for RNA-seq data}.
\newblock {\em Bioinformatics}, 30:2517--2518.

\bibitem[Soneson and Delorenzi, 2013]{Soneson2013}
Soneson, C. and Delorenzi, M. (2013).
\newblock {A comparison of methods for differential expression analysis of
  RNA-seq data}.
\newblock {\em BMC Bioinformatics}, 14.

\bibitem[Subramanian et~al., 2020]{Subramanian2020}
Subramanian, I., Verma, S., Kumar, S., Jere, A., and Anamika, K. (2020).
\newblock {Multi-omics data integration, interpretation, and its application}.
\newblock {\em Bioinformatics and Biology Insights}, 14:7--9.

\bibitem[Sun et~al., 2013]{TCCpackage}
Sun, J., Nishiyama, T., Shimizu, K., and Kadota, K. (2013).
\newblock {TCC: an R package for comparing tag count data with robust
  normalization strategies}.
\newblock {\em BMC Bioinformatics}, 14:219.

\bibitem[Tepeli et~al., 2020]{Tepeli2020}
Tepeli, Y.~I., {\"{U}}nal, A.~B., Akdemir, F.~M., and Tastan, O. (2020).
\newblock {PAMOGK: A pathway graph kernel-based multiomics approach for patient
  clustering}.
\newblock {\em Bioinformatics}, 36:5237--5246.

\bibitem[Ullmann et~al., 2022]{ullmann2022over}
Ullmann, T., Beer, A., H{\"u}nem{\"o}rder, M., Seidl, T., and Boulesteix, A.-L.
  (2022).
\newblock Over-optimistic evaluation and reporting of novel cluster algorithms:
  an illustrative study.
\newblock {\em Advances in Data Analysis and Classification}, pages 1--28.

\bibitem[Wang et~al., 2014]{wang2014similarity}
Wang, B., Mezlini, A.~M., Demir, F., Fiume, M., Tu, Z., Brudno, M.,
  Haibe-Kains, B., and Goldenberg, A. (2014).
\newblock Similarity network fusion for aggregating data types on a genomic
  scale.
\newblock {\em Nature Methods}, 11:333--337.

\bibitem[Zapf et~al., 2021]{zapf2021meta}
Zapf, A., Albert, C., Fr{\"o}mke, C., Haase, M., Hoyer, A., Jones, H.~E., and
  R{\"u}cker, G. (2021).
\newblock Meta-analysis of diagnostic accuracy studies with multiple
  thresholds: Comparison of different approaches.
\newblock {\em Biometrical Journal}, 63:699--711.

\bibitem[Zhou et~al., 2021]{Zhou2021}
Zhou, Y., Yang, B., Wang, J., Zhu, J., and Tian, G. (2021).
\newblock {A scaling-free minimum enclosing ball method to detect
  differentially expressed genes for RNA-seq data}.
\newblock {\em BMC genomics}, 22.

\end{thebibliography}

\end{document}